\documentclass[useAMS,usenatbib,12]{article}
\usepackage[dvips]{graphicx}
\title {\textbf{ Helium line detections from ELDWIM at 1.4 GHz}}
\author {Raju Baddi,$^{1}$\\  
$^1$ Raman Research Institute, C.V. Raman Avenue, Bangalore-80, India.\\ 
}
\date{}
\begin{document}

\label{firstpage}

\maketitle

\begin{abstract}
Helium line observations towards 11 Galactic positions using Westerbork Synthesis Radio Telescope(WSRT) 
have been reported. These observations were made towards nearby positions where already hydrogen lines were 
detected at sufficiently high intensity($\geq$50mK) at 1.4 GHz. This approach gave a fair chance for 
the detection of 
helium line as well, keeping in mind the relative abundance(10 \%) of helium with respect to hydrogen. 
Care was also taken to avoid the presence of HII regions along the line of sight so that the line 
emission originates from the extended diffuse low density ionized component, ELDWIM of the Galaxy. The observations 
have resulted in the detection of helium line towards 5 positions out of 11 with signal to noise ratio(snr) 
$>$ 4$\sigma$. 
An attempt has been made to associate detection/non-detection of helium line to the presence 
of surrounding HII regions. A weighting scheme that accounts for nearby($<$ 500pc) HII 
regions, their distances and other factors produces favourable results. It is seen from this weighting scheme that 
a higher weight favours the detection of helium line while lower weight is associated with  
non-detection. The idea is to correlate ionization of ELDWIM with the surrounding HII regions.
\end{abstract}

\section{Introduction}
  The presence of a diffuse Extended Low Density Warm Ionized Medium(ELDWIM) within the 
inner Galaxy has been established by low frequency($<$2GHz) Radio Recombination Line(RRL or RL for short) 
observations(Lockman 1976 ; Mezger 1978 ; Ananatharamaiah 1985). This component is understood 
to have a typical electron density $n_e$ of $1-10 cm^{-3}$ at temperatures of $10^{3}-10^{4}$ K. 
The ELDWIM was first considered by Mezger(1978) who called it as the "extended low density fully 
ionized gas" which extends from the Galactic center to 13kpc and 100pc above and below the 
Galactic plane. However continuum radiation from this particular medium was first discovered 
by Westerhout(1958) in his Dwingeloo survey, who also calculated the corresponding upper limits 
on density and total mass of ionized hydrogen in the Galaxy. Later investigators have chosen to 
call this component as Extended Low Density Warm Ionized Medium(Petuchowski \& Bennett 1993, Heiles 
1994). The recent estimated densities of ELDWIM are $1-10 cm^{-3}$(Murray \& Rahaman 2010) with 
temperatures of the order of 3000-8000K. The origin, morphology and ionization mechanism of ELDWIM 
are uncertain. ELDWIM has been thought to be as a collection of evolved HII regions(Shaver 1976) or 
perhaps it forms the outer envelopes of HII regions(Anantharamaiah 1985). In the later argument it 
has been suggested that by the sizes of HII region envelopes inferred from observations and due 
to their large number almost every line of sight intersects these envelopes in the inner Galaxy
($l<40^o$) giving the observed Galactic ridge recombination lines(Anantharamaiah 1986).
More recently(Murray \& Rahaman 2010) ELDWIM has been considered to be a diffuse gas ionized 
by massive stellar clusters unrelated to HII regions. 
The present observations aimed at detecting helium lines from ELDWIM to understand its ionization 
spectrum. The helium ionization potential(24.6 eV) is higher than that of hydrogen(13.6 eV). The 
detection and amplitude of He line would indicate the ionization condition of ELDWIM. Earlier RL 
observations have indicated that the ratio of the number of helium to hydrogen ions($N_{He+}/N_{H+}$) 
in ELDWIM is in the range 0 to 0.054(Heiles et.al 1996b). This ratio is smaller than the generally 
accepted cosmic abundance of helium, $N_{He}/N_{H} \sim $ 0.1(Poppi et.al 2007 and references therein). 
Indicating all helium in ELDWIM to be 
partially ionized. According to order of magnitude calculations(Heiles et.al 1996b; see also Domgoergen 
\& Mathis 1994) this observed ratio in ELDWIM 
requires the surface temperature of the ionizing star to be $<$35000K, if a standard HII region condition 
were to be considered. The current knowledge of initial mass function and the total Galactic star 
formation rate make it difficult to realize such a cool spectrum together with the total Galactic 
ionization requirement for the ELDWIM and HII regions. This has been called as the \emph{ionization problem}
(Heiles 1996b). However such calculations are not completely exhaustive and alternate ways to explain 
the ionization spectrum must be found.

\section[]{Observations}
 Helium RLs are much weaker than hydrogen RLs and hence longer integration time is required 
to detect them from specific directions. WSRT was used to observe hydrogen and helium RLs from 15 different 
Galactic positions. In the incoherent addition mode WSRT offers 8 IF bands, each 
with a bandwidth of 5 MHz. These were used to detect 7 Hn$\alpha$/Hen$\alpha$ RLs with n=165 to 171 and 1 Hn$\beta$ 
RL with n=208 (Table-1). The primary objective was to detect the Hen$\alpha$ line by averaging 
the spectra from the 7 bands. 
Observations were carried out using dual frequency switching with a shift of 2MHz, keeping the average 
of the hydrogen and helium rest frequencies at the center of the band. The resolution of the spectra is 
$\sim$ 4km/s. The 15 positions were constructed 
from a list of previously observed(Lockman et.al 1976, 1989; Heiles 1996a) directions which exhibited strong($\geq$50mK) 
hydrogen RLs with a single component at 1.4 GHz. A typical line strength of 50mK for hydrogen would imply 
$T_{He}$ = 0.025$\times T_{H}$ = 1.25mK for helium, taking into account the mean 0.025  
of the earlier mentioned $N_{He+}/N_{H+}$ ratio. With this strength 
for He lines the proposed integration time aimed at a 3$\sigma$ detection. Care was also taken 
to avoid occurence of HII regions along the line of sight. This was to ensure line origin from ELDWIM. 
A list of relevant transitions and the emerging rest frequencies for hydrogen 
and helium RLs is given in Table-1. \\

\begin{table}
\begin{center}
 \label{transitions}
 \begin{tabular}{ccc}
 \hline
 Transition & $\nu_{H}$ (GHz) & $\nu_{He}$ (GHz) \\
 \hline

  $165\alpha$ & 1.450716 & 1.451307 \\
  $166\alpha$ & 1.424734 & 1.425314 \\
  $167\alpha$ & 1.399368 & 1.399938 \\
  $168\alpha$ & 1.374601 & 1.375161 \\
  $169\alpha$ & 1.350414 & 1.350965 \\
  $170\alpha$ & 1.326792 & 1.327333 \\
  $171\alpha$ & 1.303718 & 1.304249 \\
  $208\beta$  & 1.44072  & \\
  
 \hline
 \end{tabular}
 \caption{List of 8 transitions and rest frequencies corresponding to each band.}
\end{center}
\end{table}     

The 7 spectra from different IF bands naturally had different velocity  
resolutions. To averge them together they were resampled and shifted using 
fourier transform interpolation. Which is essentially representing the 
spectrum by decomposed fourier components. Once the parameters of these  
components are available the spectrum could be replotted with any resolution.
With a common resolution and alignment 
the spectra were averaged by weighting each spectrum by the inverse of the 
variance in the spectrum. However it should be noted 
that the spectra corresponding to H167$\alpha$ was dropped due to the occurence 
of H210$\beta$ line near to the He167$\alpha$ line. These averaged spectra have 
been displayed in Figure 1 \& 2 after a 3-box car smoothing to improve the snr.  
The parameters obtained from gaussian fits to these spectra are given in Table-2.

\section[]{Data Analysis}
The data was acquired in fits format. Auto-correlations were extracted from the fits file using 
cfitsio library. These auto-correlations were power spectra for 2 polarizations obtained for 
1-minute data per frequency setting(LO1 \& LO2) 
designated as $T_{on}$ \& $T_{off}$, with a shift of 2 MHz. The 48 minute observation towards 
each position gave 48 spectra for each polarization. The frequency switching was done every minute. So each setting 
had 24 spectra for each antenna, 14 for WSRT. A $T_{on}/T_{off} - 1$ of these spectra removed the 
background power simultaneously correcting for the gain variation across the band. Since the primary 
goal was to detect the He line care was taken to avoid any spectrum contaminated with interference. 
A visual examination of each 1 minute integrated spectrum was done before counting it in the averaging 
group. An average of all spectra in this group was followed by an appropriate folding to further 
average the lines appearing in the two parts of the spectrum due to frequency switching. Also the two 
polarization powers were combined into one at a suitable intermediate stage. \\

As mentioned in sec 2 WSRT offers 8 IF bands in the incoherent addition mode which were used to 
detect the transitions given in Table-1. No helium line could be confidently distinguished or recognized 
in a single band averaged spectrum. The spectra from the first 7 bands except the one for H167$\alpha$ were 
averaged by fourier transform interpolation as explained in sec 2 \& the previous paragraph. A 3-box car 
smoothing was further 
applied to improve the snr of this folded 6 band averaged spectrum. Finally an average system temperature
(measured per minute) was used to calibrate the spectrum. These plots have been displayed in Figure 1 \& 2 
for 11 positions out of 15. 4 of the positions had corrupt data and had to be dropped. Gaussian parameter 
fits to the smoothed spectra have been given in Table-2.

\begin{center}
\begin{figure}[h]
\includegraphics[width=80mm,height=70mm,angle=0]{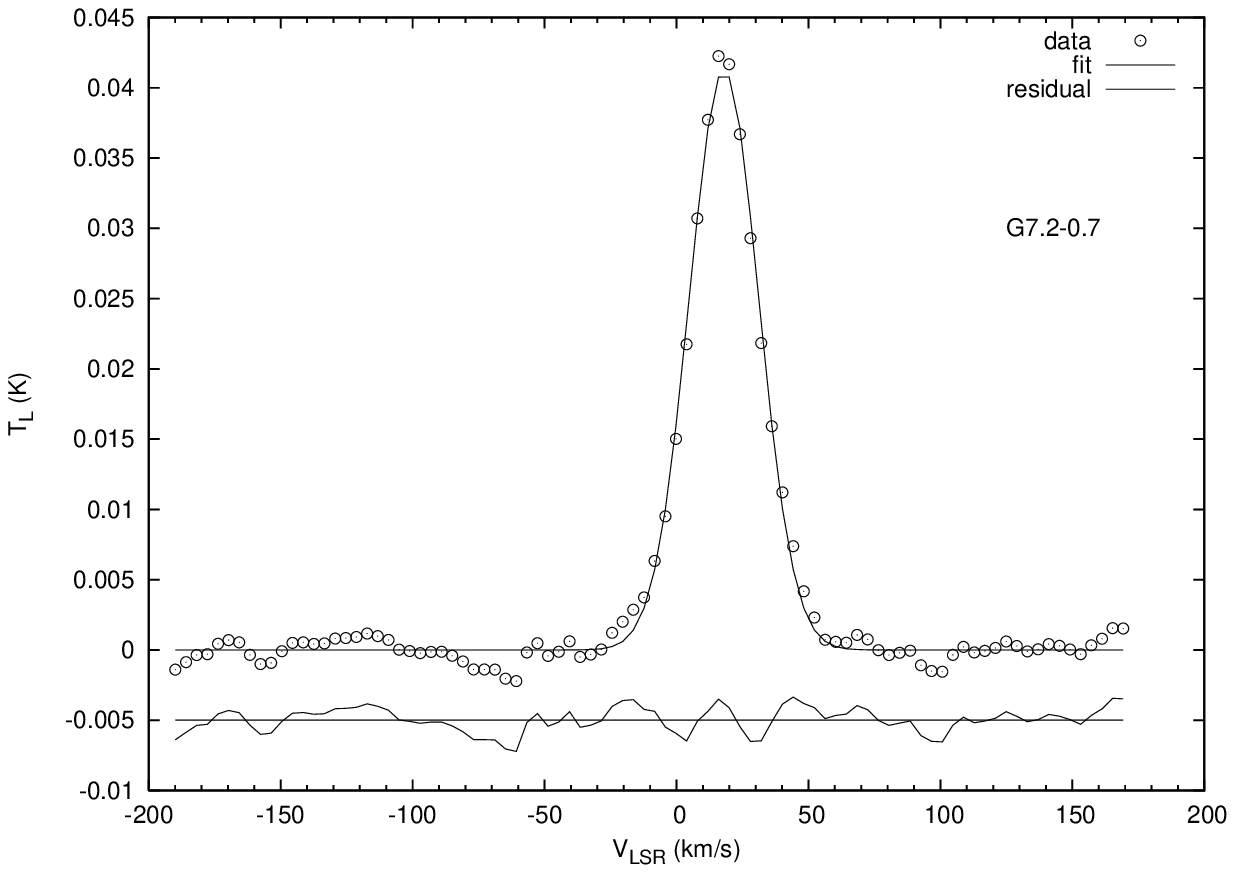}
\includegraphics[width=80mm,height=70mm,angle=0]{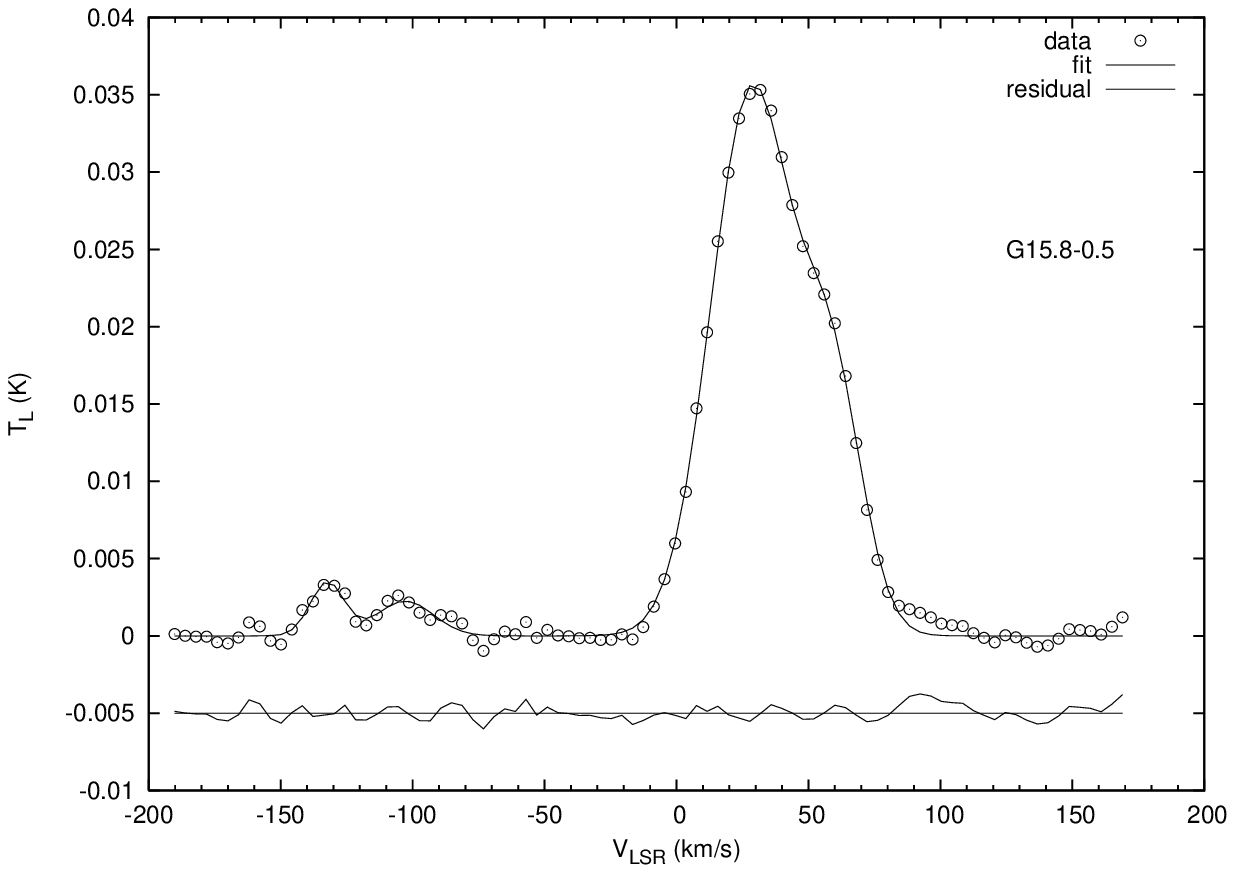}
\includegraphics[width=80mm,height=70mm,angle=0]{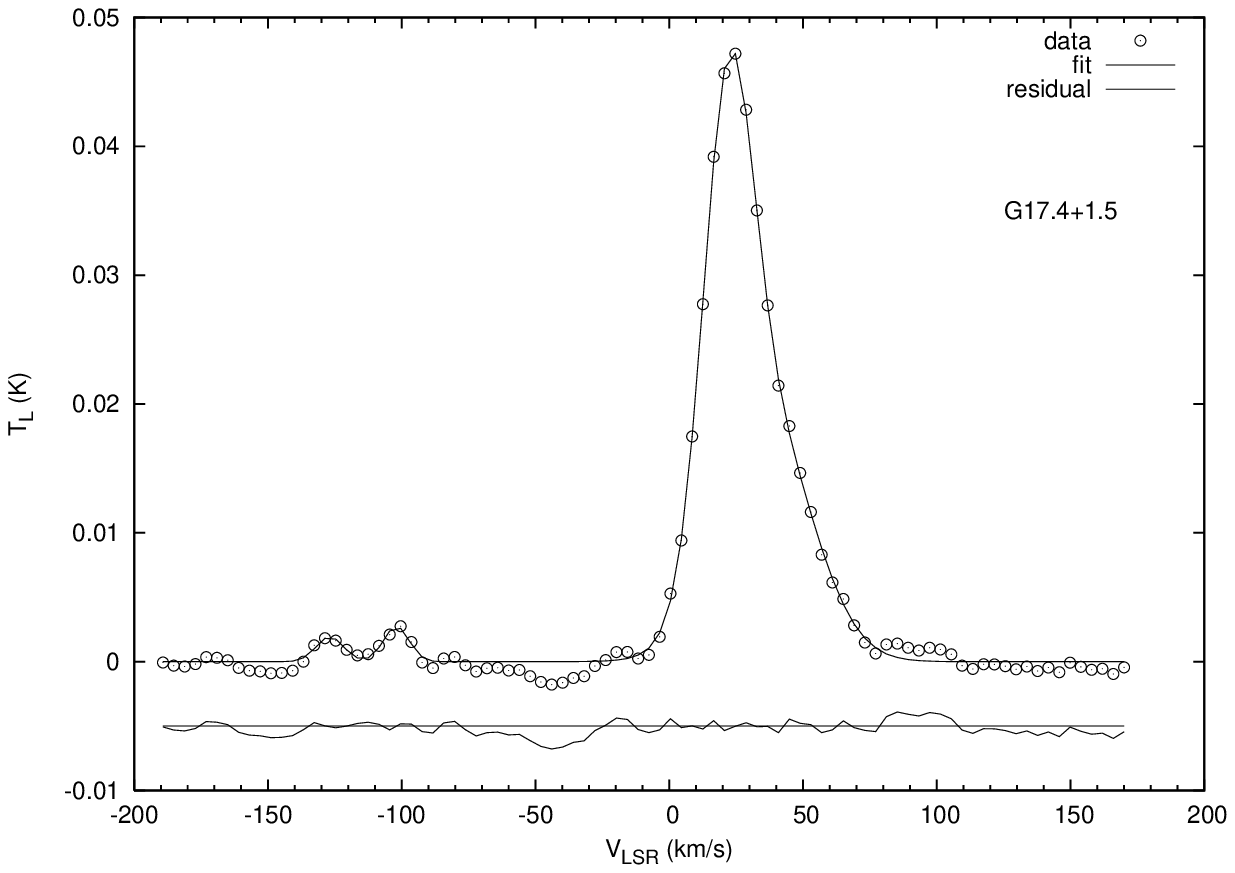}
\includegraphics[width=80mm,height=70mm,angle=0]{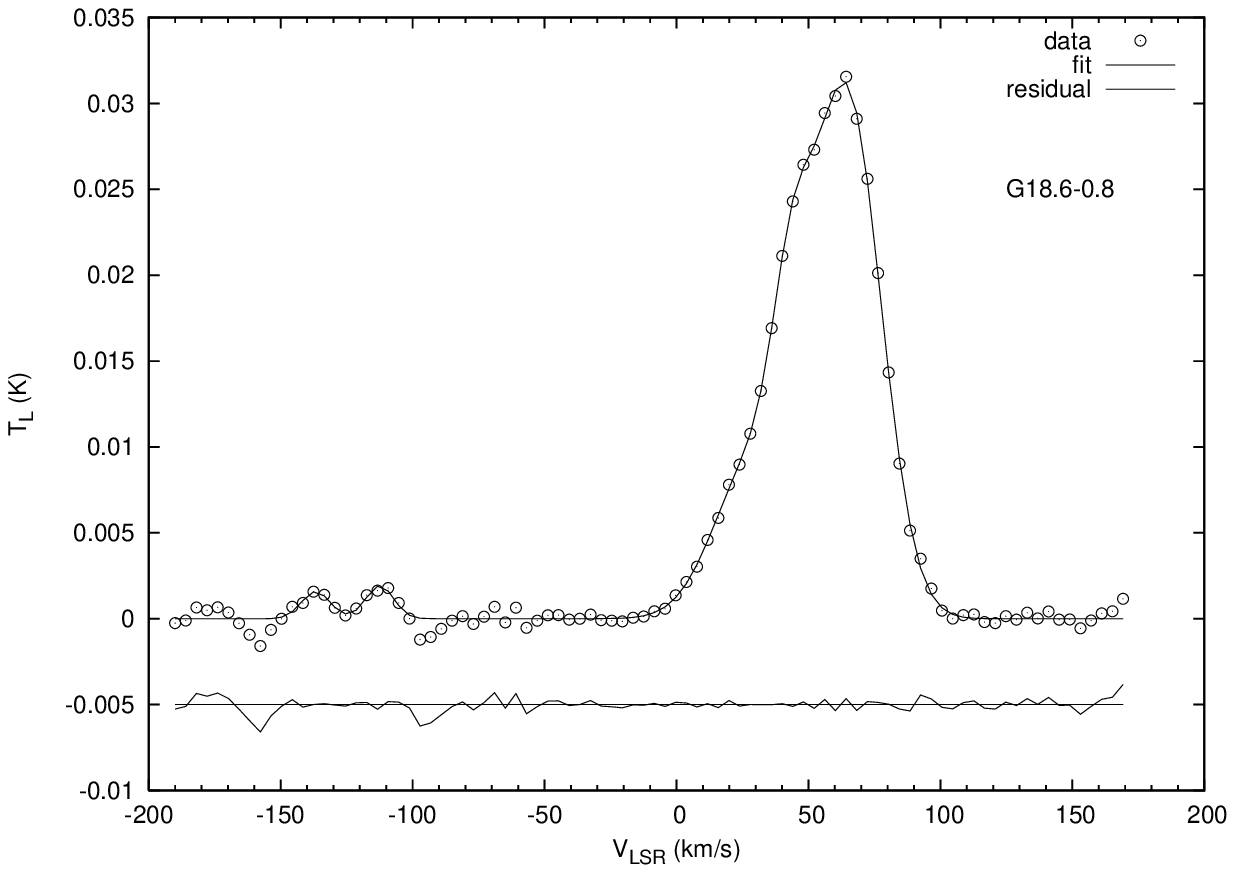}
\includegraphics[width=80mm,height=70mm,angle=0]{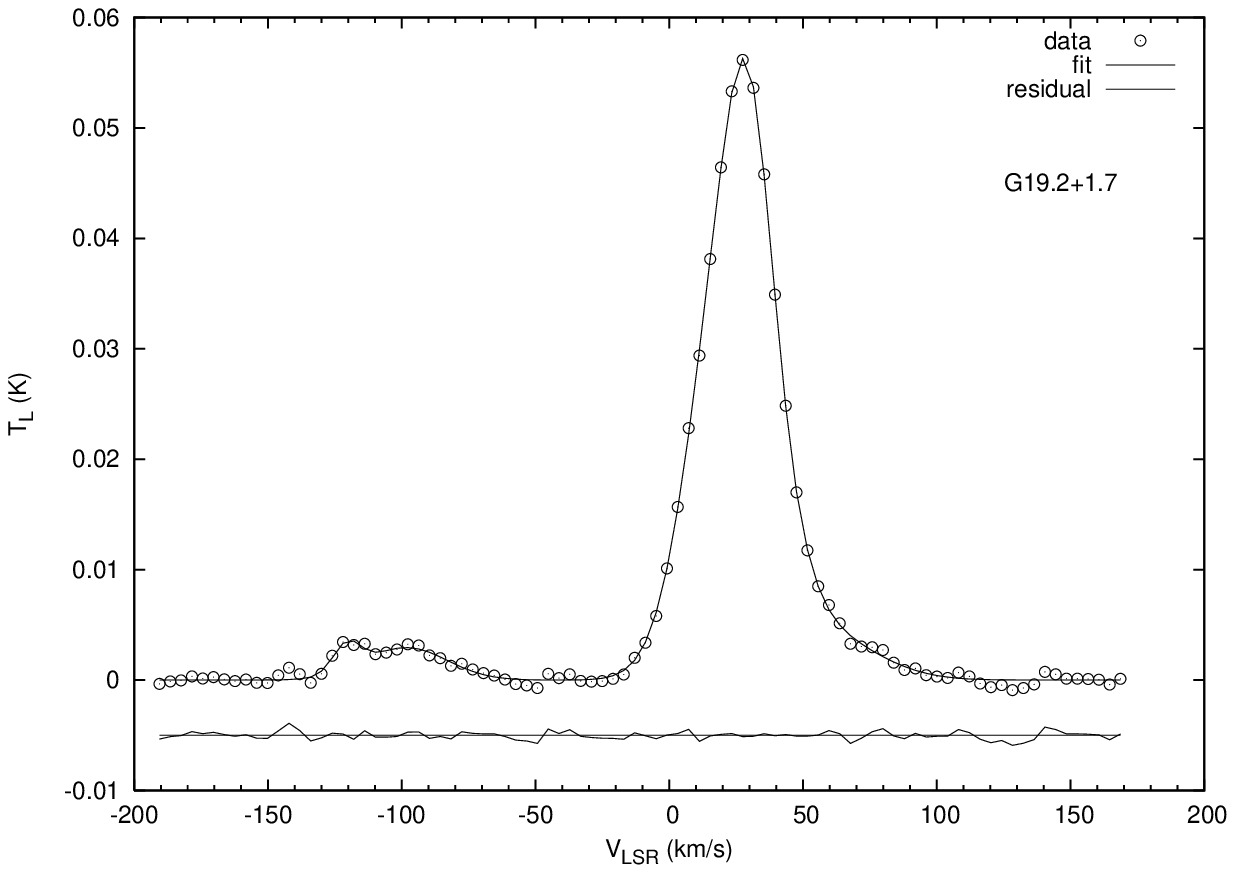}  
\hspace{4.7mm} \includegraphics[width=80mm,height=70mm,angle=0]{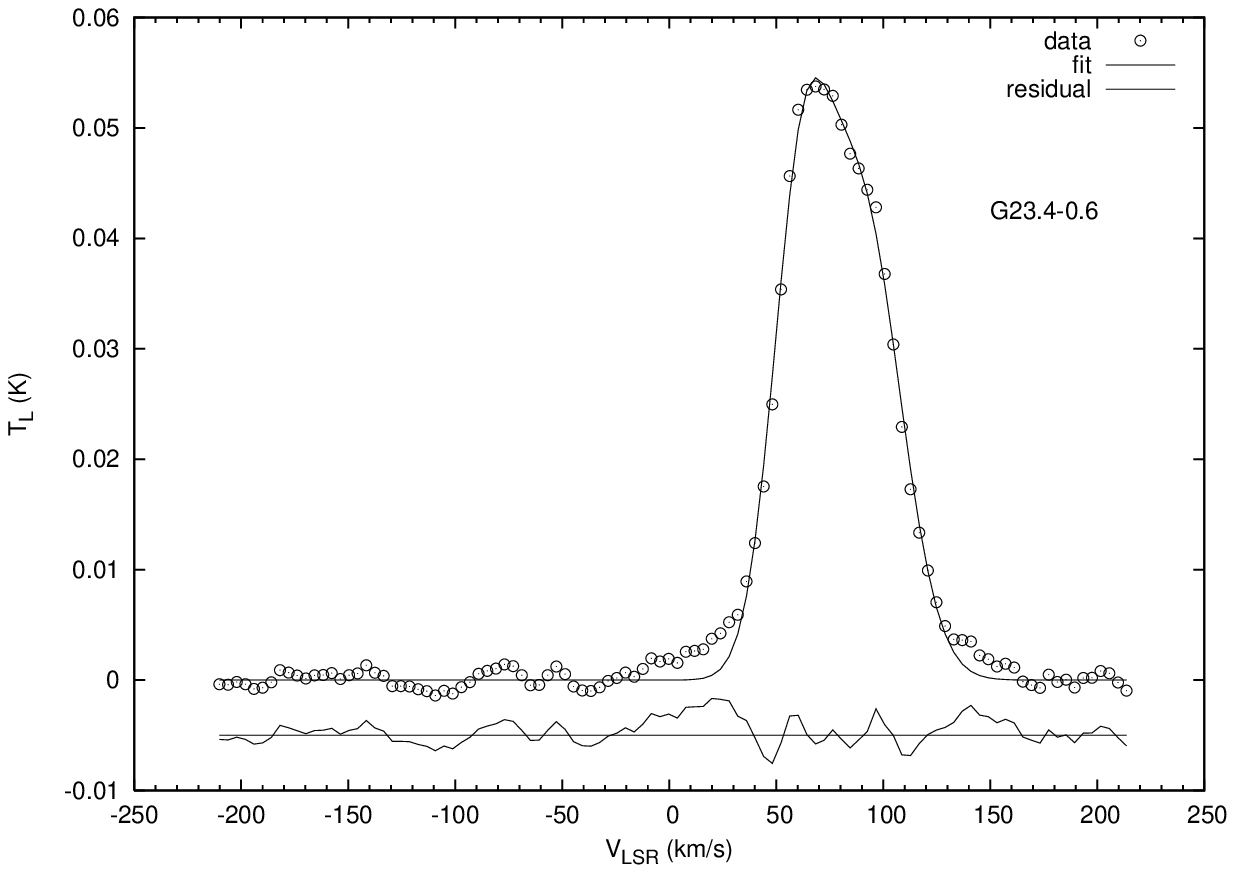}
\caption{Positions: G7.2-0.7, G15.8-0.5, G17.4+1.5, G18.6-0.8, G19.2+1.7 
G23.4-0.6. Helium line is expected at an offset of -122.2 km/s from the hydrogen line. 
The obtained gaussian parameters are given in Table-2. The residual after subtracting 
the fit from data has been shown at an offset of -0.005 along the $T_L$-axis. It should be noted that 
the final spectra 
have been corrected for poor baselines by polynomial fitting to portions of the spectrum not containing any 
astronomical spectral line. Also before polynomial fitting any residual interference was edited out to avoid 
contribution to the polynomial fit.  }
\end{figure}
\end{center}
\begin{center}
\begin{figure}[h]
\includegraphics[width=80mm,height=70mm,angle=0]{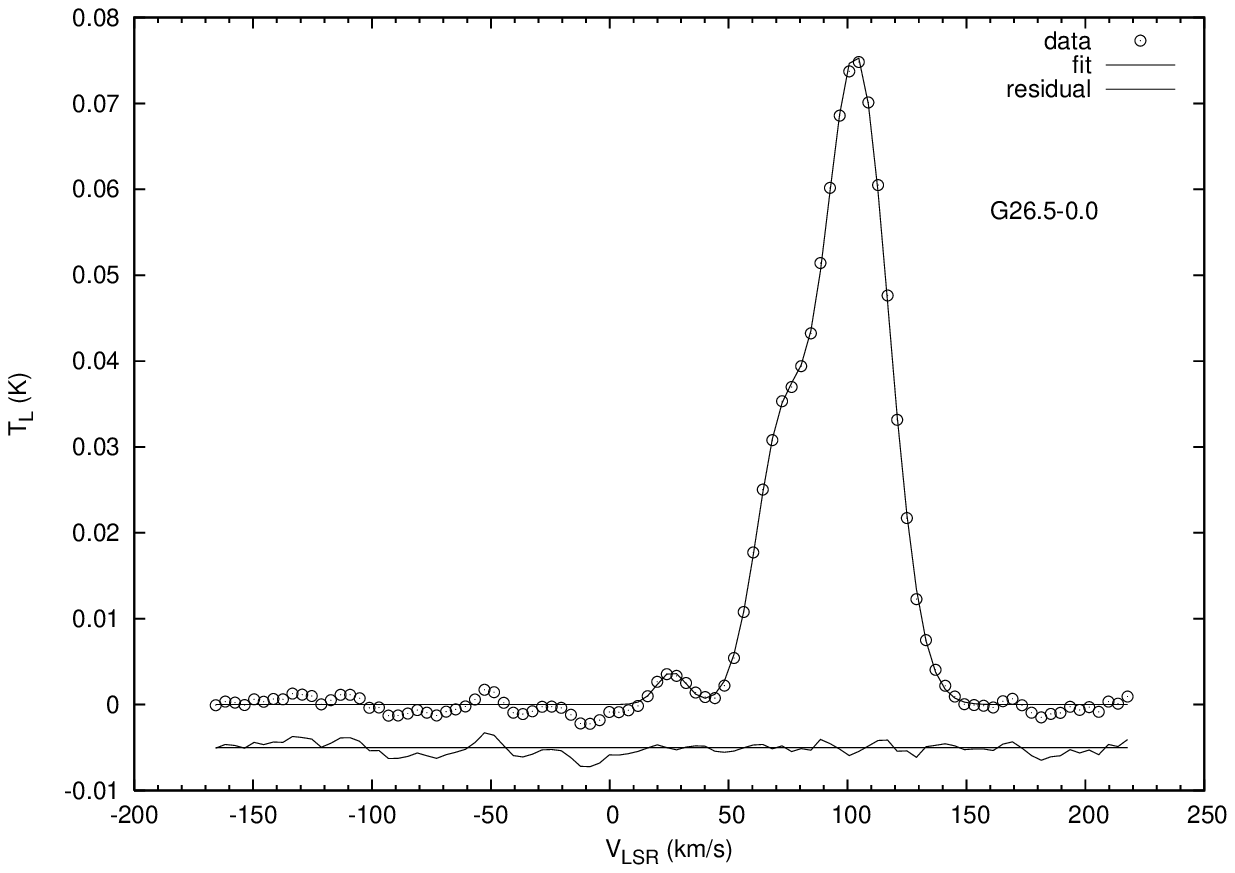}
\includegraphics[width=80mm,height=70mm,angle=0]{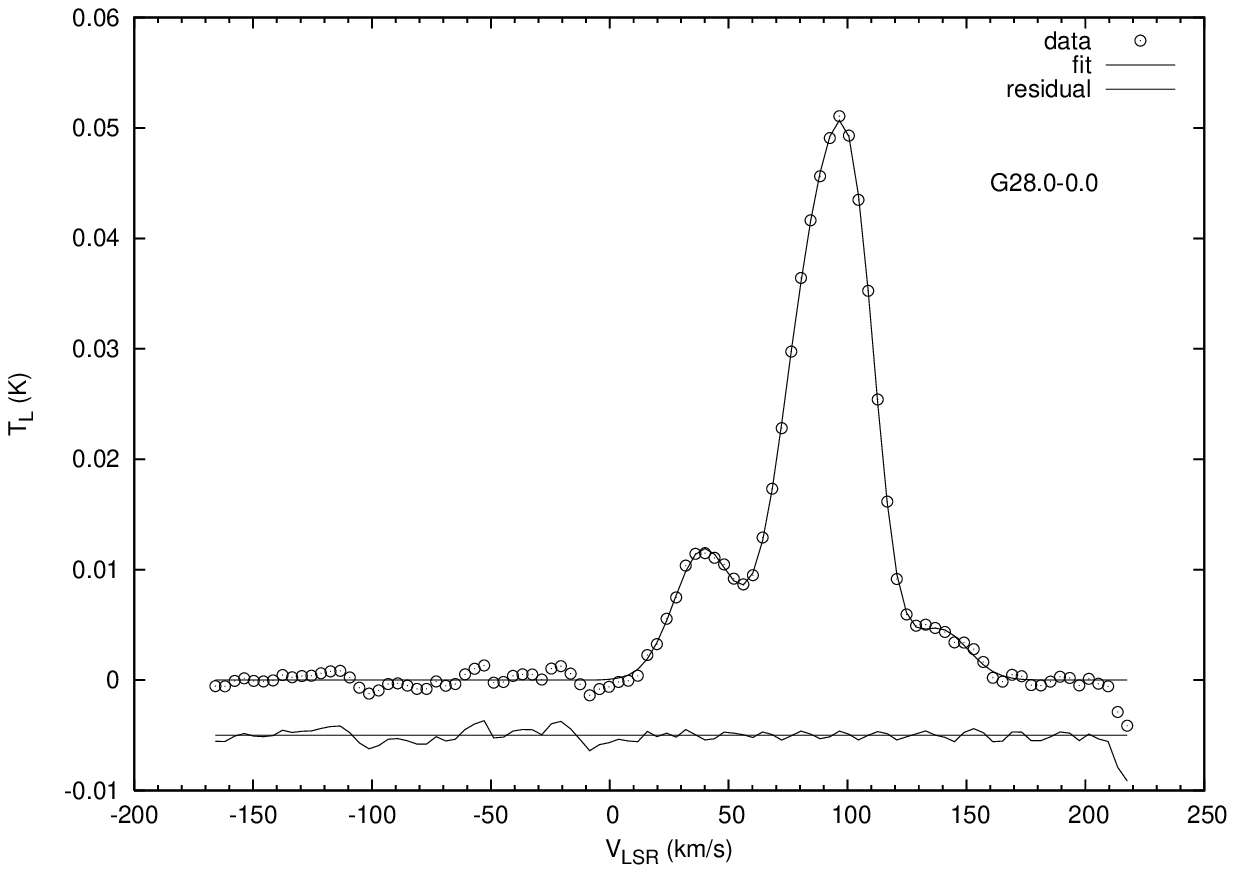}
\includegraphics[width=80mm,height=70mm,angle=0]{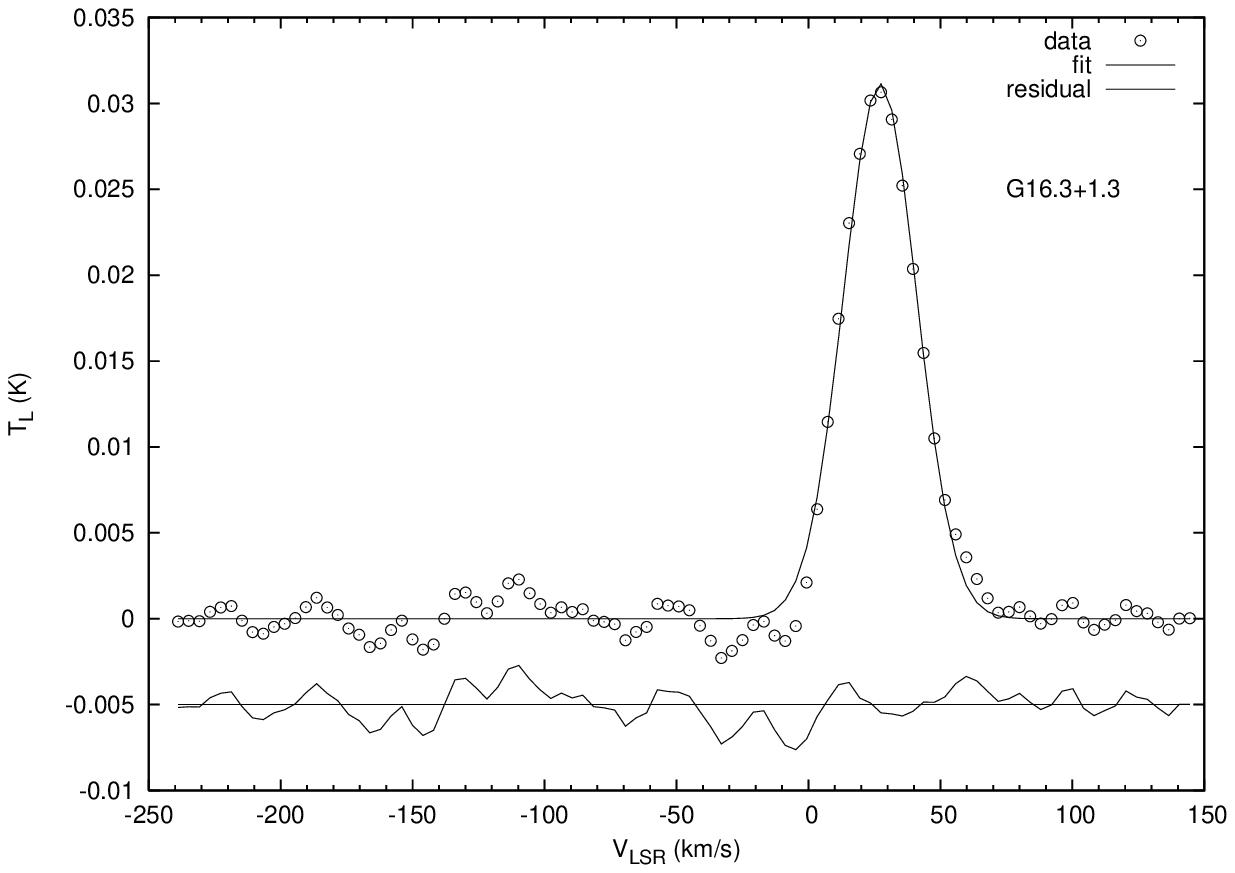}
\includegraphics[width=80mm,height=70mm,angle=0]{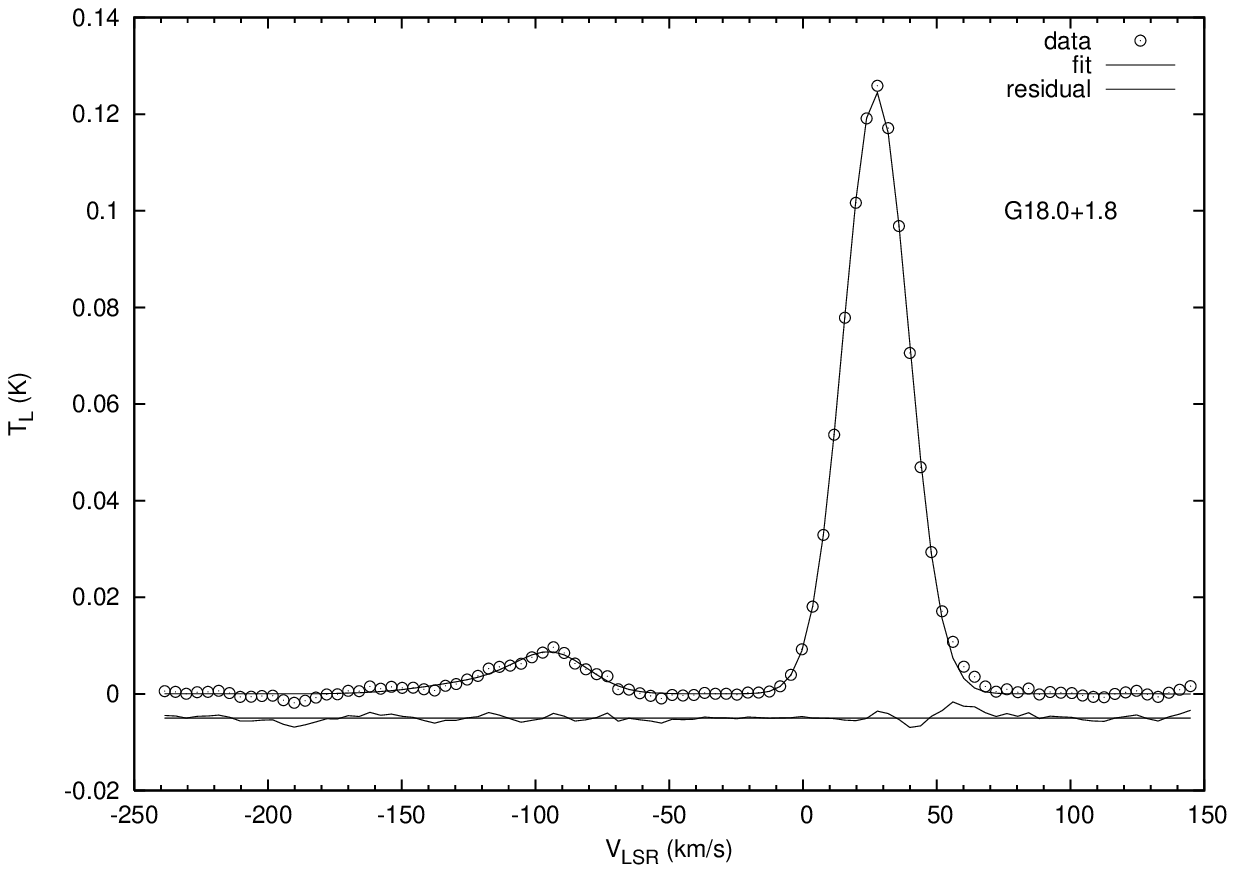}
\includegraphics[width=80mm,height=70mm,angle=0]{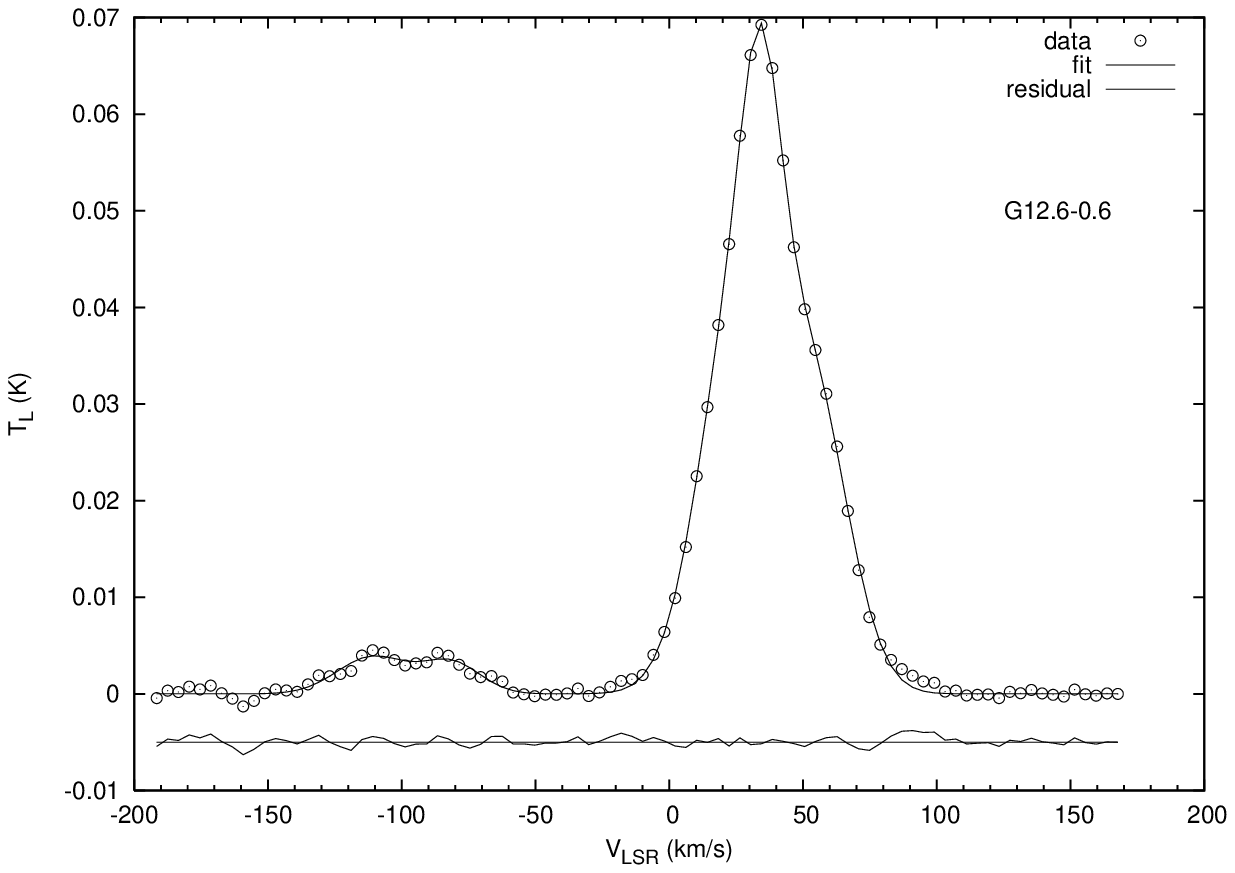}
\caption{Positions:  G26.5-0.0, G28.0-0.0, G16.3+1.3, G18.0+1.8, G12.6-0.6. 
Helium line is expected at an offset of -122.2 km/s from the hydrogen line. The obtained gaussian 
parameters are given in Table-2. The residual after subtracting 
the fit from data has been shown at an offset of -0.005 along the $T_L$-axis. It should be noted that the final spectra 
have been corrected for poor baselines by polynomial fitting to portions of the spectrum not containing any 
astronomical spectral line. Also before polynomial fitting any residual interference was edited out to avoid 
contribution to the polynomial fit.}
\end{figure}
\end{center}
\clearpage
 \begin{table}
\begin{center}
 \label{parameters}
 \begin{tabular}{cccccccccc}
 \hline
 \multicolumn{2}{c}{Source} & \multicolumn{2}{c}{$T_{L}$ (mK)} & \multicolumn{2}{c}{$V_{LSR}$ (km/s)} 
& \multicolumn{2}{c}{$ \Delta V_{LSR}$ (km/s)} & {$\sigma$} &  \\ \cline{1-8}
   $l^{o}$ & $b^{o}$ & H & He & H & He & H & He & (mK) & {$N_{He+}/N_{H+}$} \\
 & & & & & & & & & \\ \hline
 & & & & & & & & & \\
  7.2&-0.7&41.2(0.5)&-&18.0(0.2)&-&31.0(0.4)&-&0.74&$<$0.018 \\
 & & & & & & & & & \\
 & & & & & & & & & \\
 15.8&-0.5&35.2(0.4)&2.25(0.3)&28.4(0.5)&-102.8(2.0)&36.3(0.8)&25.2(5.0)&&0.064(0.01) \\
 & & 15.4(0.8) & & 59.0(0.8) & & 27.1(1.2) & & 0.41 & \\
 & & & & & & & & & \\
& & & & & & & & & \\
 17.4&+1.5&33.4(2.2)&2.73(0.6)&21.9(0.2)&-102.0(1.2)&21.7(0.9)&11.2(2.8)&& 0.082(0.02) \\
 & & 18.7(1.6)&&35.7(1.7)&&40.8(1.4)&&0.66& \\
 & & & & & & & & & \\
& & & & & & & & & \\
 18.6&-0.8&9.1(1.8)&-&29.3(6.1)&-&34.8(6.7)&-&&  \\
 && 10.3(3.8)&&43.7(0.8)&&17.7(2.8)&&0.69& \\
 &&30.4(0.1)&&64.5(0.7)&&30.5(1.0&&&$<$0.023 \\
 & & & & & & & & & \\
& & & & & & & & & \\
 19.2&+1.7&41.8(6.2)&2.9(0.2)&24.1(1.5)&-98.2(2.6)&34.6(1.1)&35.7(5.4)&&0.069(0.015) \\
 &&14.8(3.5)&&30.4(0.4)&&18.8(1.8)&&0.41&\\
 &&4.2(2.1)&&55.4(18.8)&&48.6(18.6)&&& \\
 & & & & & & & & & \\
& & & & & & & & & \\
 23.4&-0.6&40.0(5.4)&-&61.4(1.5)&-&32.1(1.7)&-&& \\
 &&41.4(2.8)&&90.3(2.6)&&42.3(3.1)&&0.7&$<$0.017\\
 & & & & & & & & & \\
& & & & & & & & & \\
 26.5&0.0&3.8(0.6)&-&26.1(1.2)&-&14.9(2.9)&-&& \\
 &&29.3(0.7)&&71.6(0.47)&&25.2(0.85)&&0.78& \\
 &&75.4(0.5)&&103.4(0.2)&&32.3(0.44)&&&$<$0.01 \\
 & & & & & & & & & \\
& & & & & & & & & \\
 28.0&0.0&11.7(0.5)&-&39.4(1.1)&-&29.0(2.3)&-&&\\
 &&39.0(12.0)&&87.5(7.6)&&34.5(8.2)&&0.55& \\
 &&24.1(24.4)&&103.7(1.7)&&22.8(5.4)&&& \\
 &&4.5(0.6)&&138.7(2.7)&&28.0(5.1)&&&$<$0.014 \\
 & & & & & & & & & \\
& & & & & & & & & \\
 16.3&+1.3&31.1(0.5)&-&27.1(0.25)&-&32.6(0.6)&-&0.92&$<$0.03 \\
 & & & & & & & & & \\
& & & & & & & & & \\
 18.0&+1.8&124.5(0.5)&8.4(1.2)&27.4(0.06)&-92.8(3.5)&28.4(0.1)&26.0(5.2)&0.75&0.067(0.01) \\
 & & & & & & & & & \\
& & & & & & & & & \\
 12.6&-0.6&56.1(13.0)&3.2(0.7)&34.2(1.0)&-81.5(4.2)&22.9(2.2)&25.5(6.9)&&0.057(0.026) \\
 &&22.7(8.0)&&16.9(5.5)&&27.6(4.7)&&0.46& \\
 &&28.8(2.5&&56.2(1.9)&&27.9(2.1)&&& \\
 & & & & & & & & & \\
 \hline
 \end{tabular}
 \caption{Gaussian parameters fitted to the spectra in Figure 1 \& 2. The values in the paranthesis are errors. Regions 
without helium line detection have upper limits defined as $\sigma/T_{L}(H)$. All the ratios have been calculated 
assuming the helium line to be associated with the strongest hydrogen component. It should be noted that the associated 
carbon line parameters have not be tabulated. Carbon RL is expected at an offset of $\sim-150 km/s$ w.r.t 
the hydrogen RL.}
\end{center}
\end{table}
\clearpage
\section[]{Discussion of parameters in Table-2}
 	Table-2 shows that out of 11, 5 positions(G15.8-0.5,G17.4+1.5,G19.2+1.7,G18.0+1.8 \& G12.6-0.6) 
exhibit He line detections with snr $>$ 4$\sigma$. For regions 
were hydrogen and helium are homogeneously ionized the ratio $N_{He+}/N_{H+}$ is equal to the ratio of observed line 
strengths of He and H. In case of non-detection of He line an upper limit was obtained by taking 
the ratio of rms of noise to the strongest H line strength. The highest ratio is towards G17.4+1.5 with a value of 0.082. 
But this value comes from the assumption that the helium and hydrogen zones overlap. 
Some spectra exhibit pressure broadening, clearly visible with the H component having extended wings at the 
edge of the profile. Especially the spectra to be mentioned are G23.4-0.6 and G19.2+1.7. The non-detection 
or lower intensity of He line than expected implies that it is not ionized or is partially ionized in ELDWIM. 
Density bound HII regions can act as source of hot ionizing photons(Ferguson et.al 1996; Wood \& Mathis 2004).  
The escaping photons will ionize ELDWIM in the vicinity. However the photon spectrum will be 
drastically modified(Osterbrock 1989; Hoopes \& Walterbos 2003; Wood \& Mathis 2004) by the gas surrounding 
the HII regions. Since any photon that can ionize helium can also ionize 
hydrogen(Osterbrock, 1989). Only those higher energy photons that escape hydrogen, which is much more abundant than 
helium, can ionize helium in ELDWIM. This investigation has revealed a high value of $N_{He+}/N_{H+}$ ratio $\sim$
0.082 which is close to the primordial cosmic abundance $\sim$0.1(Poppi et al 2007 and references therein). \\

There are also interesting positions like G28.0+0.0 \& G26.5+0.0 in the Galactic plane which show a lack of He 
line signature. More observations or a further study of observations towards these and surrounding regions can reveal 
more about the ionization of ELDWIM in these directions. It is required to have an understanding of distribution of 
HII regions towards these directions. A non-detection indicates that either helium is under abundant or is not ionized 
within these regions. The detection of helium line can also be reverted back to the morphology and ionization 
spectrum originating from the HII regions. Lack of He line indicates that the photons leaking from the surrounding 
HII regions have a cooler spectrum whereas directions towards which He line is seen have a stronger ionizing spectrum.
This investigation has produced 5 positions towards which distinct He line profiles have been observed.
Indicating clearly the presence of diffuse ionized helium. RL detections towards nearby HII regions by earlier 
observers(Heiles et.al 1996a,b) match closely with the $V_{LSR}$ of the current observed lines, but however are stronger. \\

The line widths of H and He lines is another interesting aspect. Towards some regions like G17.4+1.5 \& 
G15.8-0.5 the line width of He is nearly half of H width, within error bars. This is expected in the case of pure thermal 
doppler broadening. However turbulence can make He line width more than half of H line width. 
The width agreement between He and H is a criterion to hold the two lines to be originating 
from the same region. In the case(G17.4+1.5) where He line width is simply half of H line width it suggests that the lines 
originate in a low density medium where negligible pressure broadening \& turbulence is expected. Pressure broadening could 
contribute to the width of the He line in denser regions and produce extended wings(Smirnov et al 1984). There seems 
to be significant amount of pressure broadening towards positions like, G23.4-0.6 \& G19.2+1.7. The non-agreement 
of He and H line widths indicates that they can originate in different regions. In some cases the width of He line is 
larger than H line this may be due to blending of different components into one. Further the constraint of observed 
$V_{LSR}$ difference($\sim$122.2 km/s) between H and He suggests that the region of origin moves at the same velocity along the line of 
sight. Based on differential Galactic rotation is the same cloud or a nearby cloud. During gaussian fitting of 
parameters it has been assumed that the He line is primarily associated with the strongest H line. This seems to be 
true towards positions G17.4+1.5, G19.2+1.7 \& G18.0+1.8 where the $V_{LSR}$ for He and H are in good agreement 
within the error bars. Broad He lines are seen towards positions which have multiple H line components. It was not 
possible to fit similar multiple components to He line and obtain consistent parameters with H line widths \& amplitudes. 
In case of G18.0+1.8 \& G19.2+1.7 a complete agreement of $V_{LSR}$ between He and H is seen but the 
widths do not seem to be only thermal. The extra width of He indicates turbulence.\\ 

Another interesting region in the sample is 
G18.6-0.8 which exhibits multiple H line components. Even though line like features are seen towards $V_{LSR}$ of 
-112 km/s, the fitted parameters are not consistent with the expected $V_{LSR}$ difference between He and H lines to be 
originating from the same region. This position seems to cover an HII region at the edge within the beam(Figure 3 \& 4). 
It is likely that the prominent H \& He lines come from different regions for this direction. The set of positions with 
no He line detection are G7.2-0.7,G23.4-0.6,G26.5+0.0,G28.0+0.0 \& G16.3+1.3. The 
position G7.2-0.7 shows a strong H line but has no associated He line which is expected from the fitted parameters 
to H line at $V_{LSR}$ of -104 km/s. However this position is a good candidate for inspecting ELDWIM and its ionization. G23.4-0.6 
is another good candidate for ELDWIM, perhaps with a smaller telescope beam.  
The H line feature for this position seems to show significant pressure broadening, the most distinct in the sample. This position 
accepts 2 Voigt profiles instead of 2 Gaussians.  
This may indicate contribution from denser regions. G16.3+1.3 due to its poor snr may still be a valuable candidate 
for He line detection. The weighting scheme(sec 5) predicts a He line detection for this position. \\

The possible kinematical distances(Sofue et.al 2009) of ELDWIM clouds due to $V_{LSR}$ of the lines for all 
the 11 positions in Table 2 have been marked in Figure 5 . The identified HII regions(Paladini et.al 2003 ; Anderson et.al 2011) 
distributed in Galactic longitude and latitude within a distance of 500pc from the line originating region have also been marked 
on the same plot to show their relative location to the ELDWIM clouds. \\
\section[]{Correlation with HII regions}
The obtained values of $N_{He+}/N_{H+}$ ratio for different positions have been tabulated in Table 2 \& 3. 
This ratio with the normal accepted primordial abundance of helium to hydrogen is expected to be 0.1(Peimbert et.al 1988; 
Osterbrock 1989; Baldwin et.al 1991; Reynolds 
1995; Heiles et.al 1996b; Madsen et.al 2006; Poppi et.al 2007). Here in most of the cases it has turned out to be smaller than this. 
When helium optical lines(Reynolds 1995, Madsen et.al 2006) or RL(Heiles et.al 1996b) are observed in HII regions this depression 
in the ratio is attributed to the smallness of the ionized He zone compared to the H zone which engulfs the former
(Osterbrock 1989). However in the case of ELDWIM gas is ionized externally, i.e ionization happens from periphery 
to inwards. In a diffuse extended gas it is expected that helium and hydrogen are ionized similarly with no 
morphological difference between them. Assuming the ionizing radiation is strong. Under such a situation the 
ratio $N_{He+}/N_{H+}$ should reveal the abundance 
ratio between the two elements assuming complete ionization. However in ELDWIM this is not seen. This ratio is less 
than the expected abundance ratio. This implies either helium is under abundant or the radiation ionizing the extended 
diffuse gas is not strong enough to ionize it completely. \\

In this paper an attempt has been made to correlate the identified HII regions(Paladini et.al 2003 ; Anderson et.al 2011) around the 
line originating region with helium RL detection. It is seen from this correlation that the ratio $N_{He+}/N_{H+}$ depends on the 
distribution of HII regions around them(Figure 5) together with the distance to the cloud from the solar system. This correlation scheme 
of ionization of ELDWIM with surrounding HII regions has been quantified as a weight W associated with each observed region given by,

\begin{equation}
 W~=~\frac{N_{HII} {D_c}^{2}}{{D_c}^{3}{D_c}^{3}}\left[\frac{1}{D_{1HII}^2}+\frac{1}{D_{2HII}^2} + \cdot \cdot \cdot +\frac{1}{D_{nHII}^2} \right]~=~\frac{N_{HII}}{{D_c}^{4}}\sum_{n}{\frac{1}{D_{nHII}^2}}.
\end{equation}

The weight has been considered to be proportional to the inverse square of the distance to the HII region $D_{nHII}$, as 
the flux from a source goes as inverse square of the distance. Since every HII region may or may not be a source of 
ionizing photons the weight has been taken to be proportional to the number of HII regions, $N_{HII}$. The distance to the cloud 
determines the amount of ELDWIM within the beam. Considering a sphere centered at the line originating region the volume 
of ELDWIM within the beam is proportional to $D_c^3$ and so is the amount of helium \& hydrogen. The photons that can ionize 
helium can also ionize hydrogen. For a given number of photons that can ionize helium, helium ionization is diluted due 
to the presence of hydrogen. The detection weight has been taken to be inversly proportional to the net amount of helium and 
hydrogen in view of a weight for the ratio $N_{He+}/N_{H+}$. This contributes $D_c^{6}$ in the denominator. 
The surface area of this volume ($\propto $$D_c^2$) measures the photon input into it. This contributes a $D_c^2$ in the 
numerator. The net contribution from all the above has been accounted by introducing a $4^{th}$ power of distance to the 
cloud in the denominator on right hand side of Equation (1).The identified HII regions have been marked around the line 
originating regions in Figure 5 as per their 
nearest kinematical distances(Sofue et.al 2009). It can be seen from Figure 5 that regions lying beyond 4 kpc from the solar 
system do not exhibit He line detection. A plot of observed ratio of $N_{He+}/N_{H+}$ Vs corresponding weight 
$Log_{10}[W_{near}]$for potential candidates representing clean ELDWIM has been given in Figure 6. It can be seen from this 
plot that higher ratio of $N_{He+}/N_{H+}$ is seen for regions with higher weight, i.e they are grouped towards the right upper 
corner of the plot. While lower weight regions are grouped towards the left lower corner of the plot. The weights towards 
different positions have been tabulated in Table-3.

\begin{figure}[h]
\begin{center}
\includegraphics[width=150mm,height=210mm,angle=0]{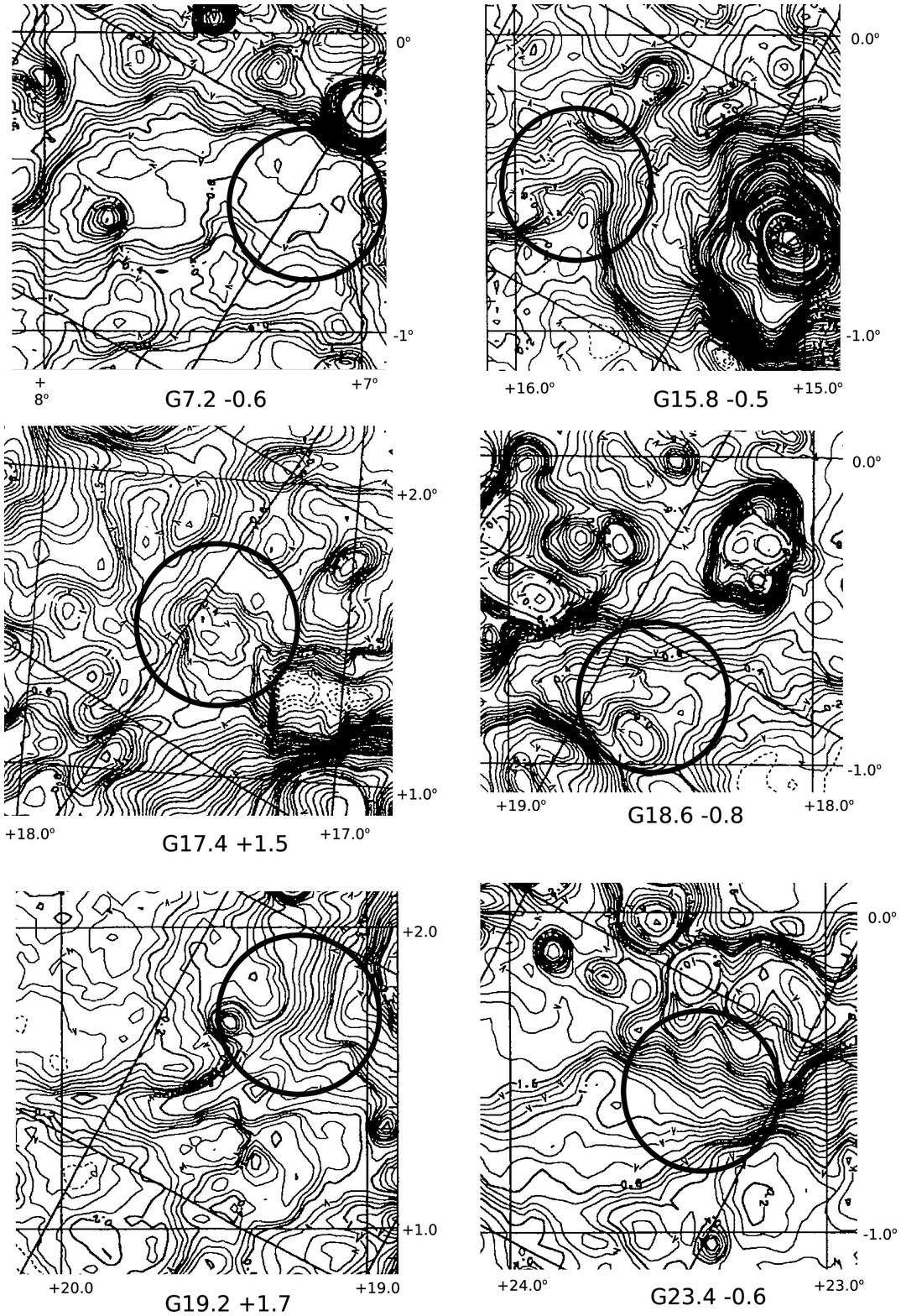}
 \caption{Positions: WSRT beam($\sim0.5^{o}$) on the first 6 of the 11 positions on 11 cm continuum map(Reich et.al 1990).}
\end{center}
\end{figure}
\begin{center}
\begin{figure}[h]
\includegraphics[width=150mm,height=220mm,angle=0]{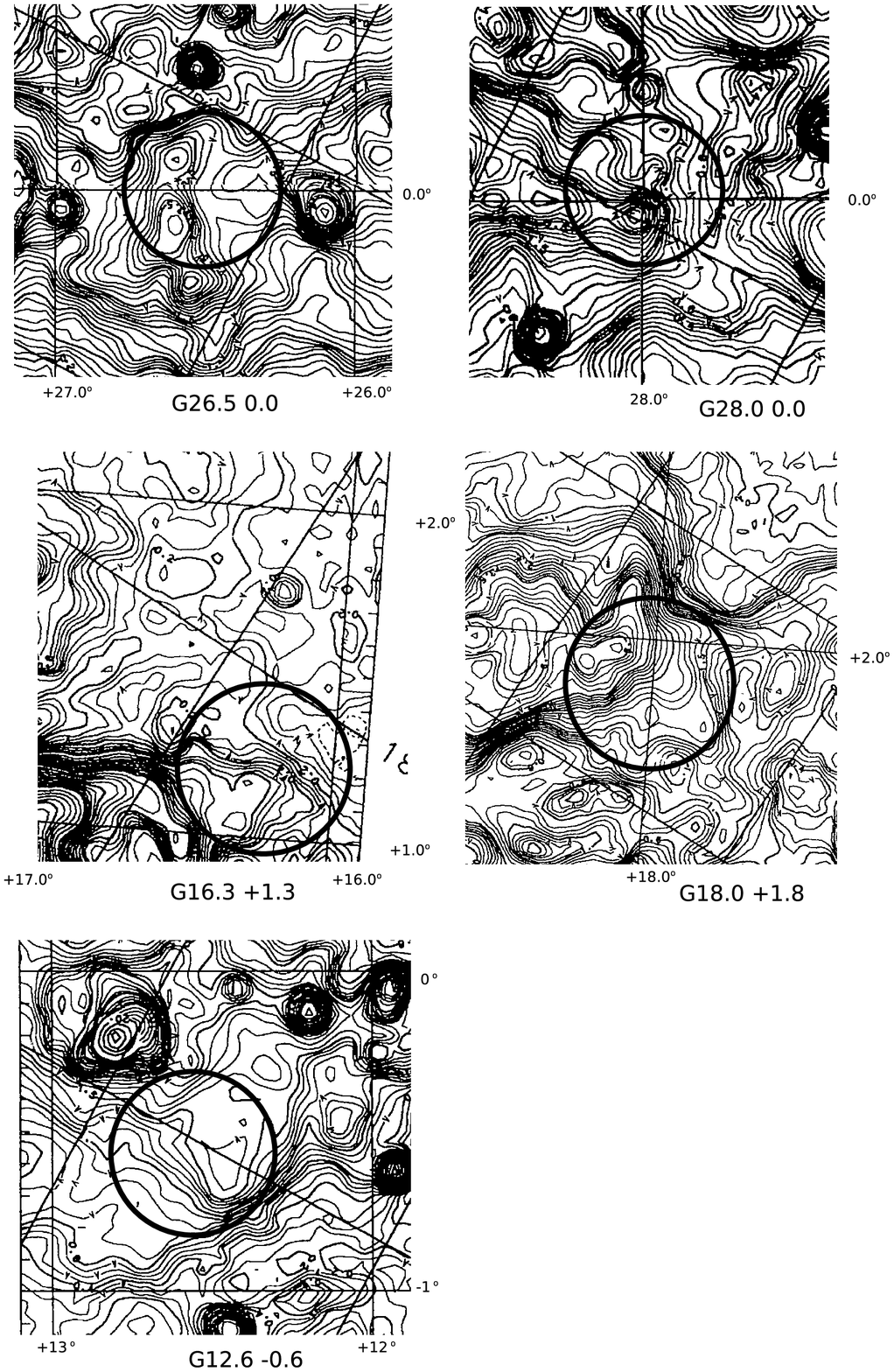}
\caption{Positions: WSRT beam($\sim0.5^{o}$) on the last 5 of the 11 positions on 11 cm continuum map(Reich et.al 1990).}
\end{figure}
\end{center}
\begin{center}
\begin{figure}[h]
\includegraphics[trim = 1mm 1mm 1mm 120mm, clip, width=63mm,angle=-90]{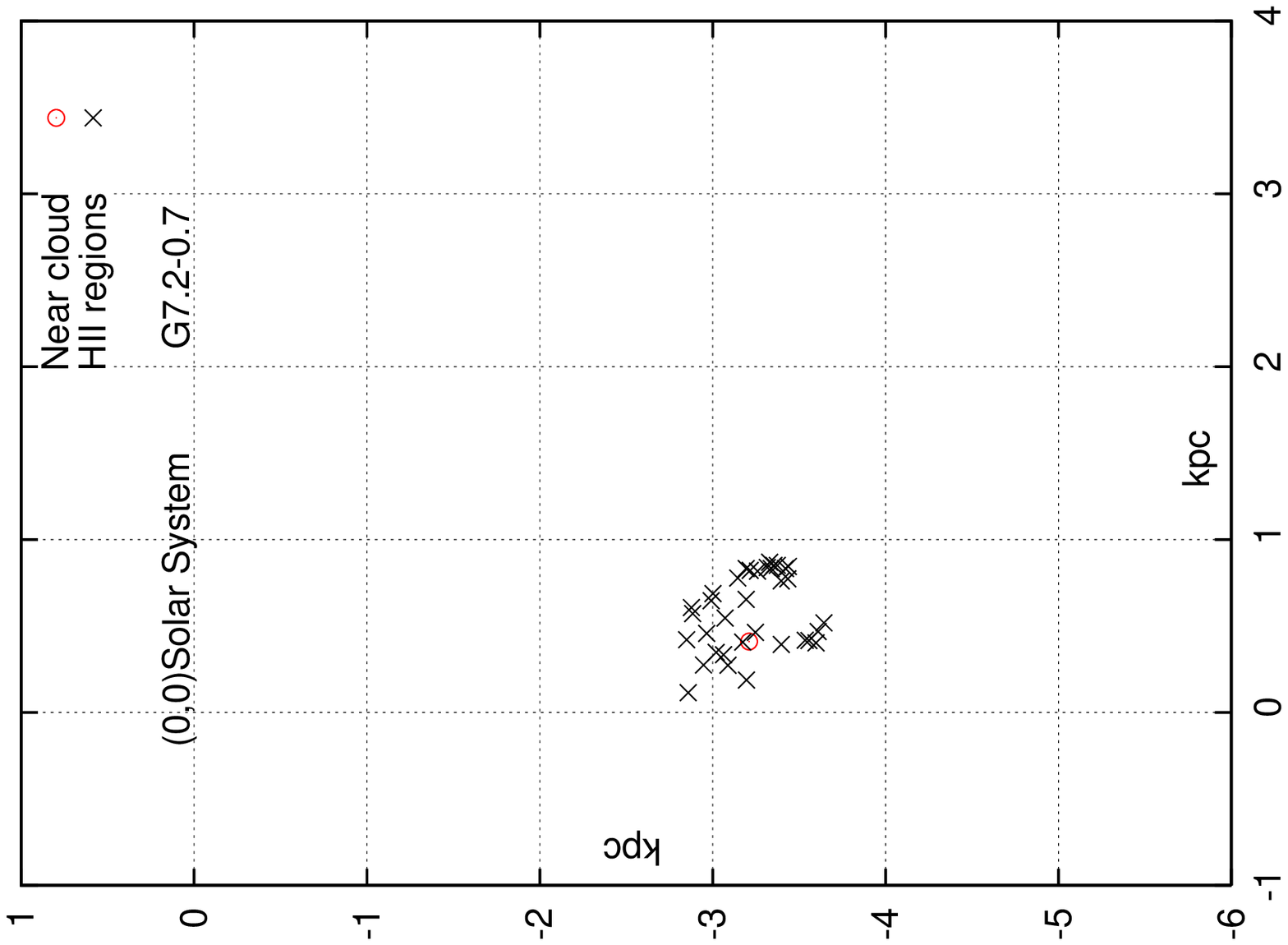}
\includegraphics[trim = 1mm 1mm 1mm 120mm, clip, width=63mm,angle=-90]{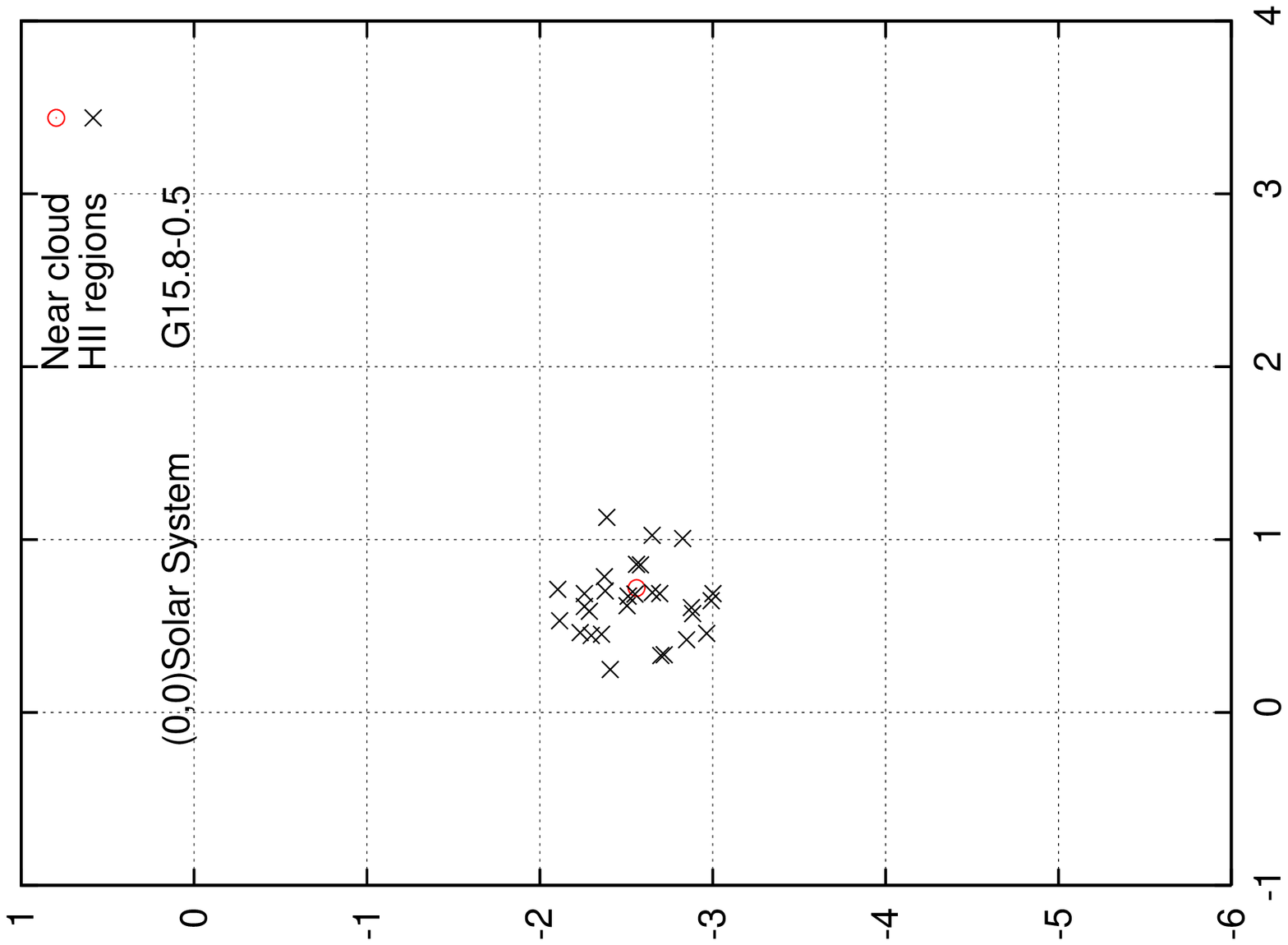}
\includegraphics[trim = 1mm 1mm 1mm 120mm, clip, width=63mm,angle=-90]{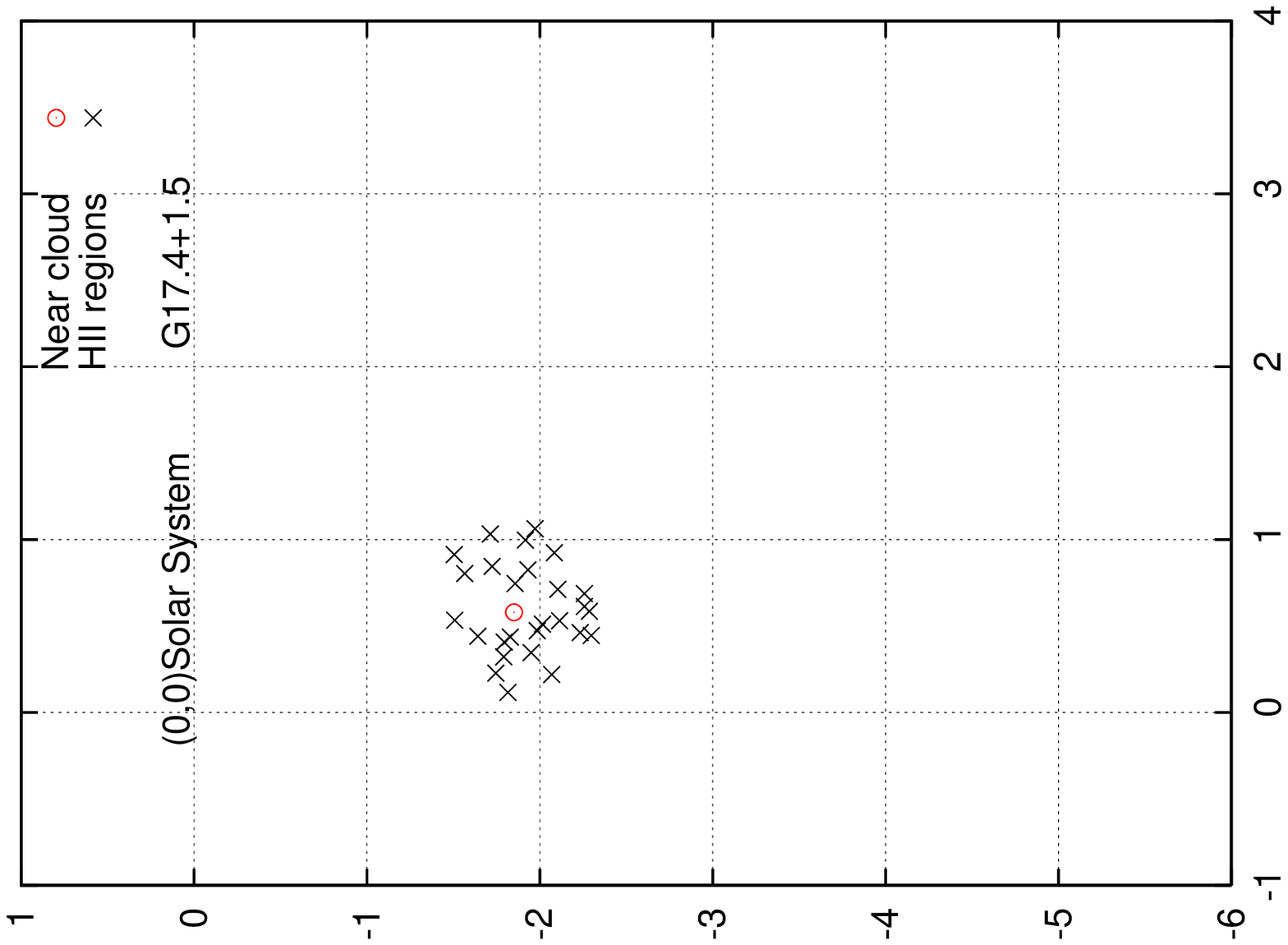}
\includegraphics[trim = 1mm 1mm 1mm 120mm, clip, width=63mm,angle=-90]{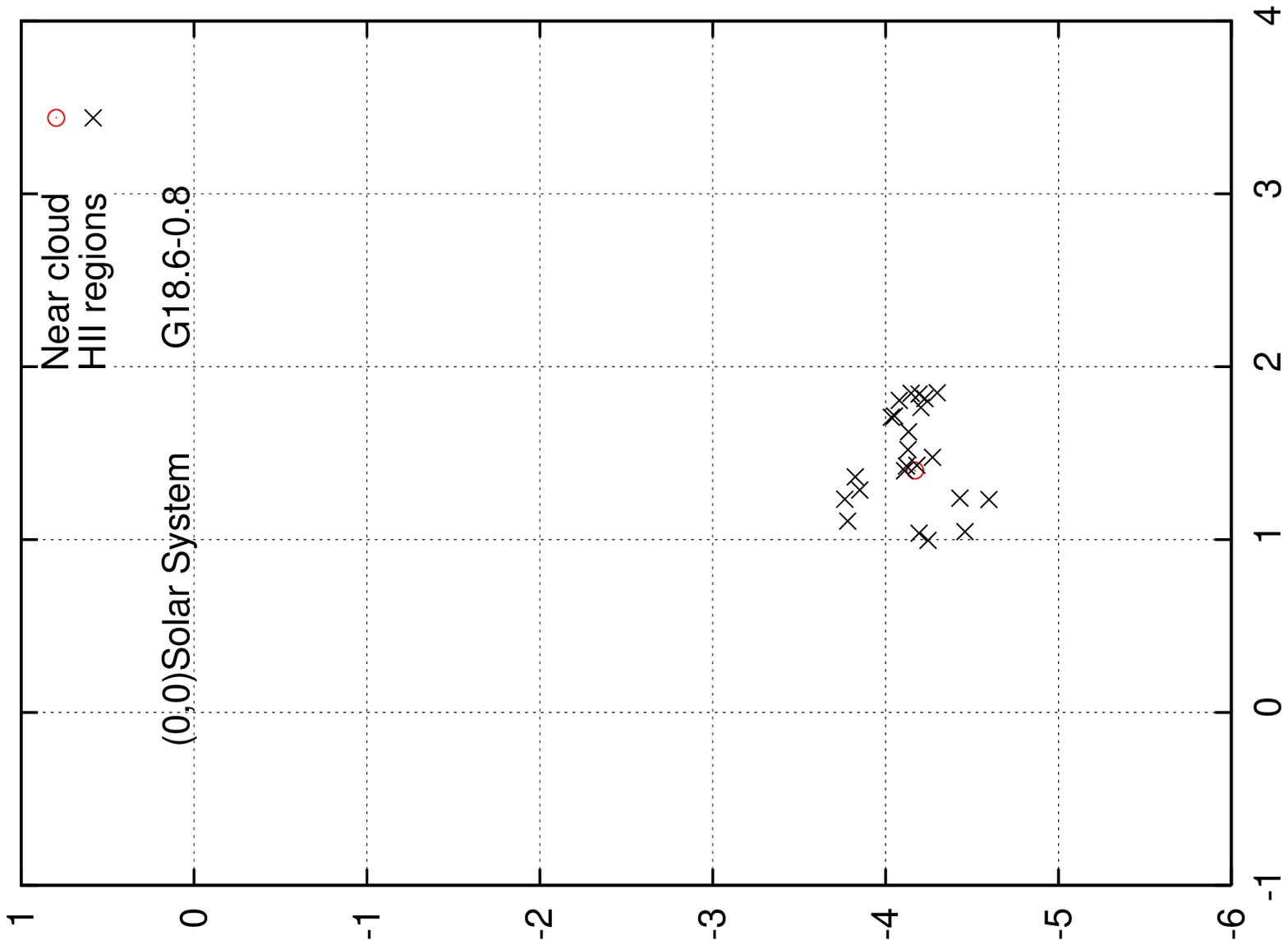}
\includegraphics[trim = 1mm 1mm 1mm 120mm, clip, width=63mm,angle=-90]{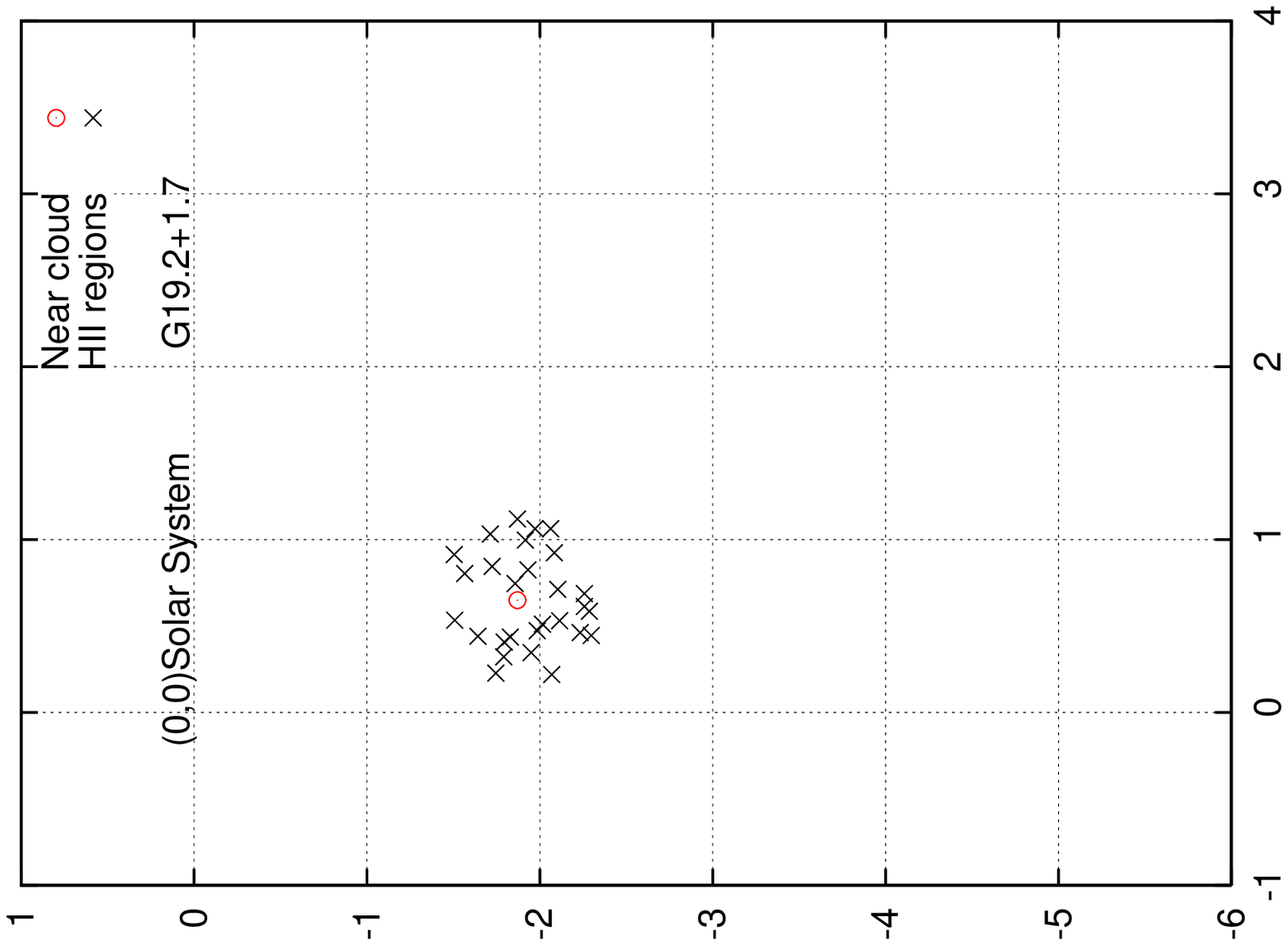}
\includegraphics[trim = 1mm 1mm 1mm 120mm, clip, width=63mm,angle=-90]{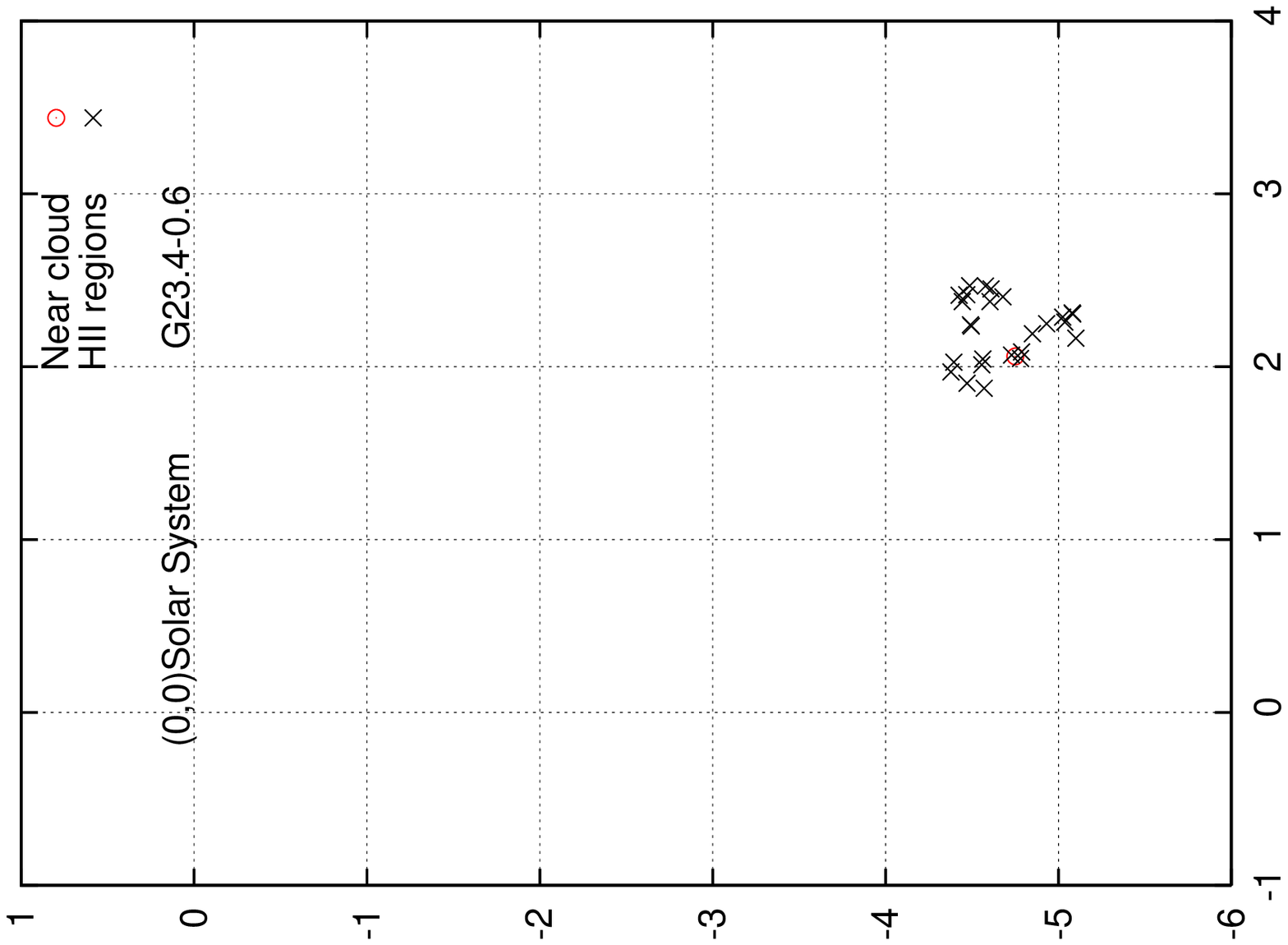}
\includegraphics[trim = 1mm 1mm 1mm 120mm, clip, width=63mm,angle=-90]{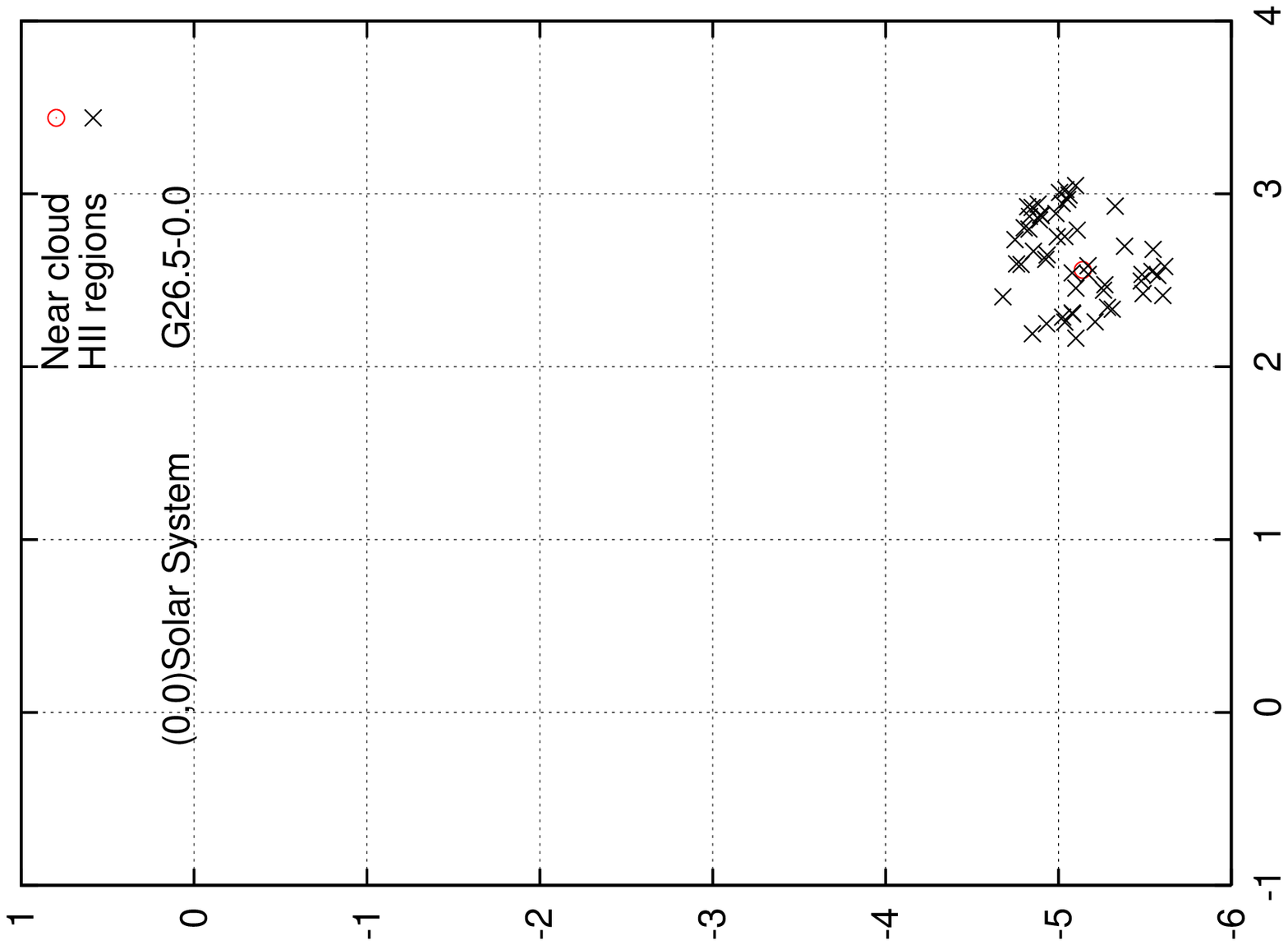}
\includegraphics[trim = 1mm 1mm 1mm 120mm, clip, width=63mm,angle=-90]{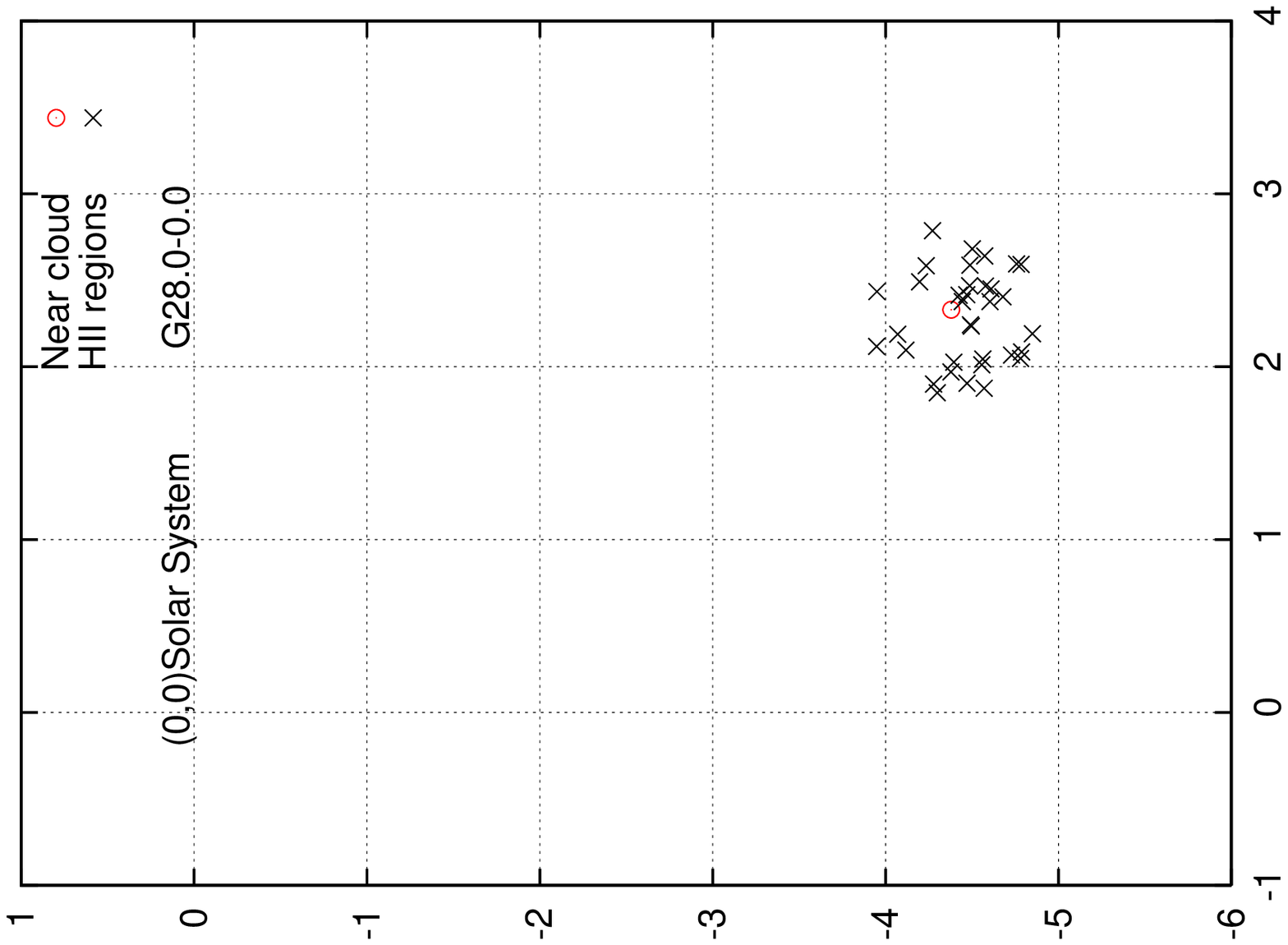}
\includegraphics[trim = 1mm 1mm 1mm 120mm, clip, width=63mm,angle=-90]{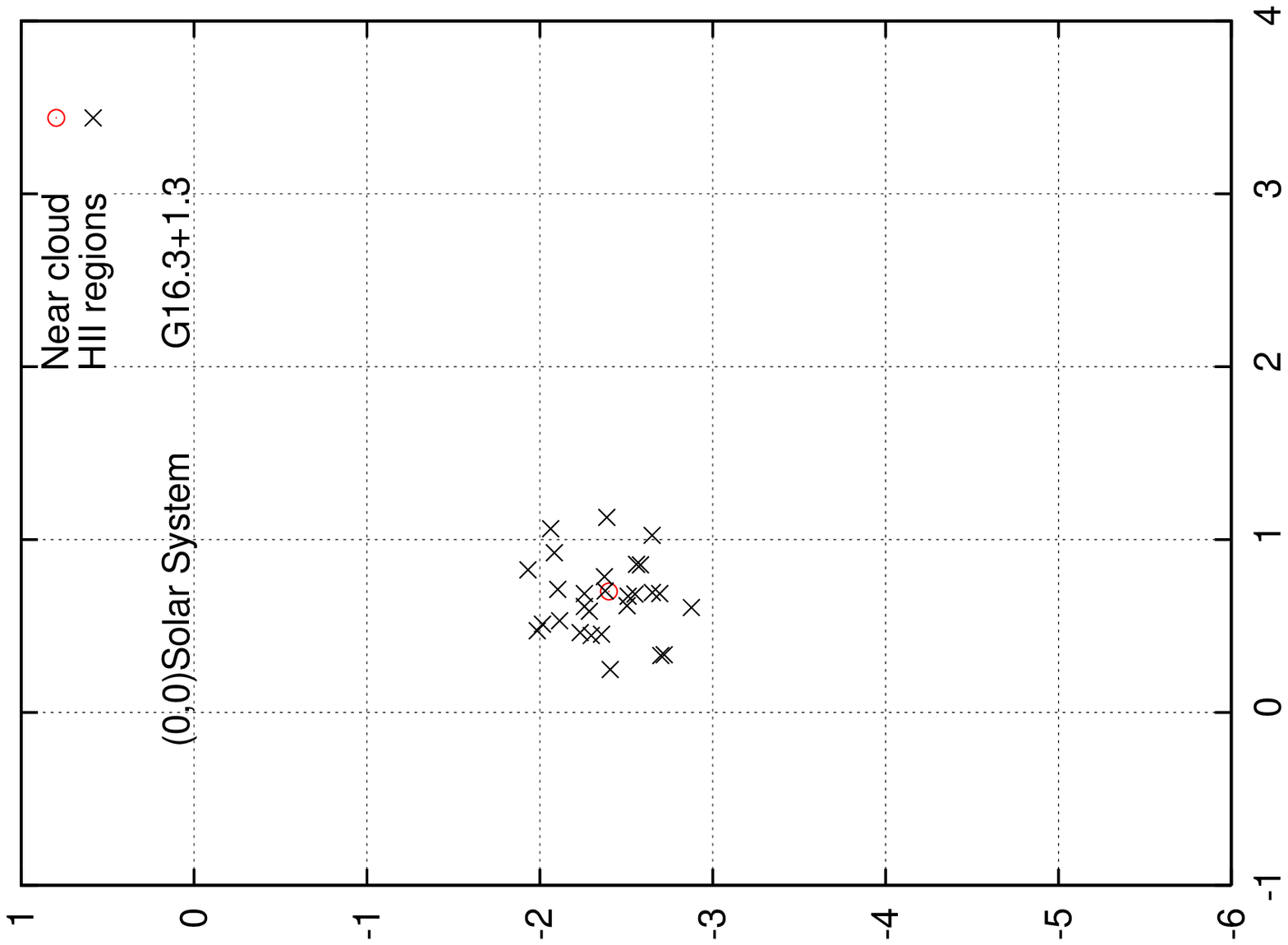}
\includegraphics[trim = 1mm 1mm 1mm 120mm, clip, width=63mm,angle=-90]{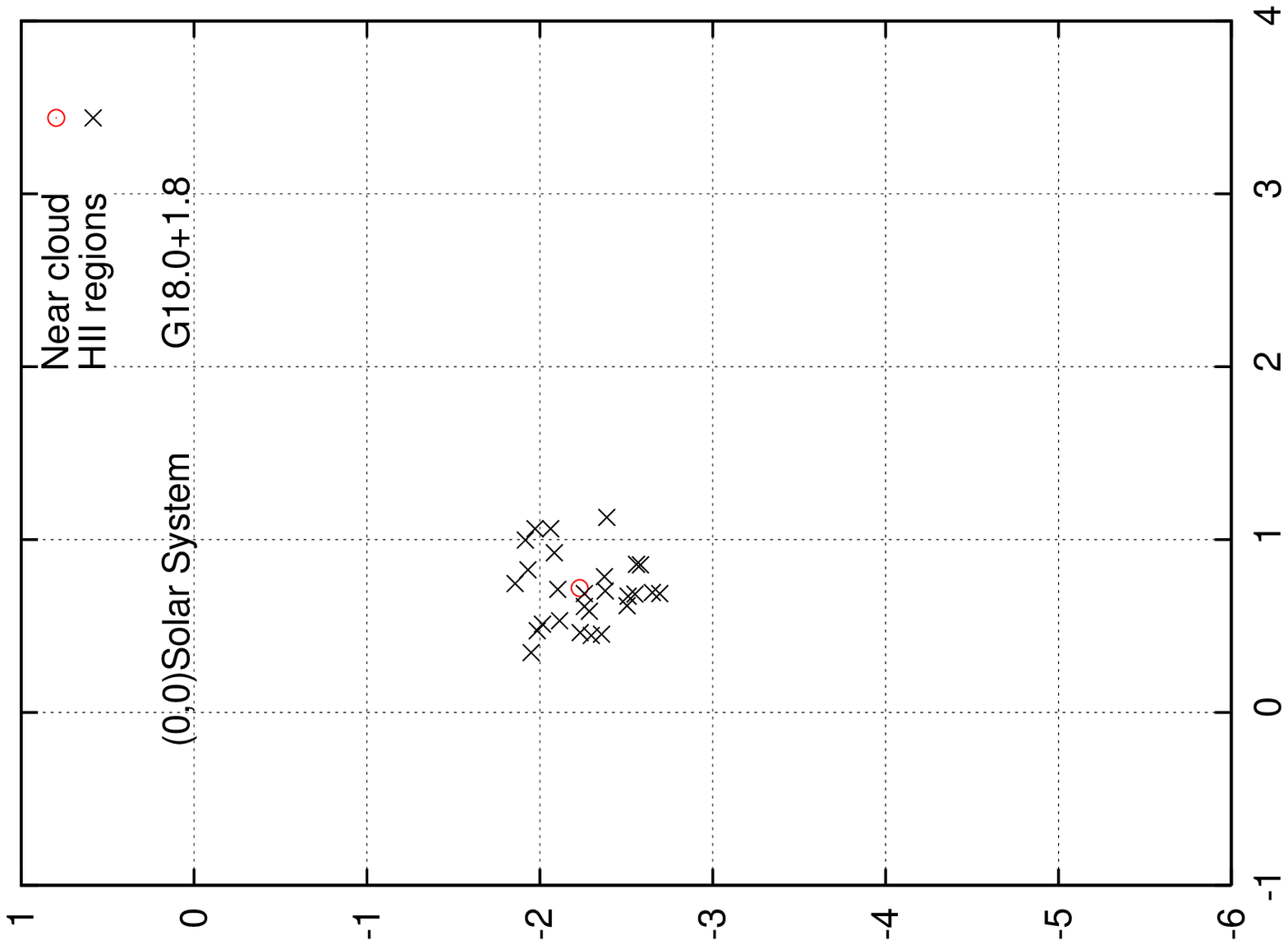}
\hspace{9mm}\includegraphics[trim = 1mm 1mm 1mm 120mm, clip, width=65mm,angle=-90]{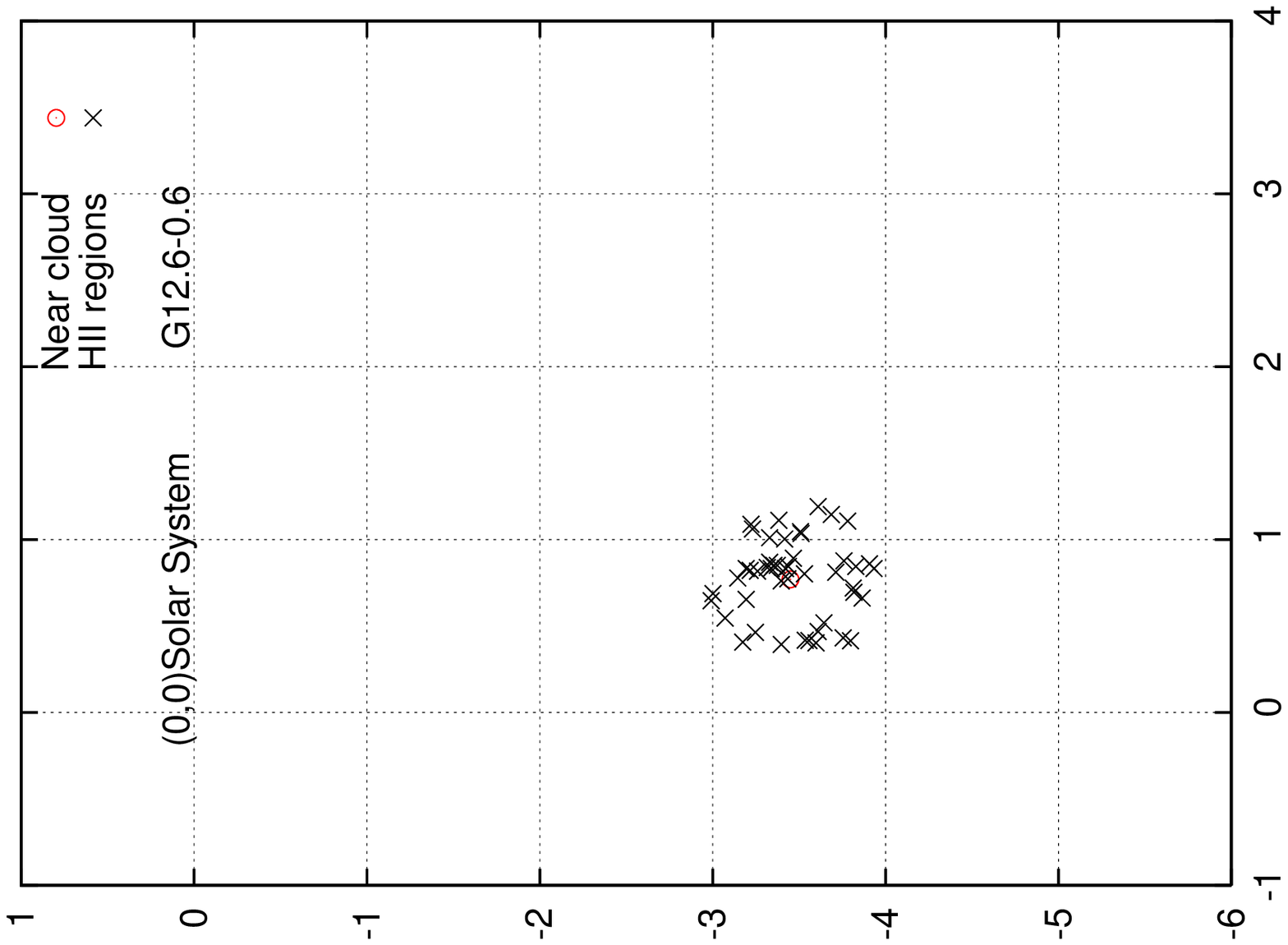}
\caption{Distribution of HII regions(Paladini et.al 2003 ; Anderson et.al 2011) around the line originating ELDWIM clouds. 
Kinematical distances are due to Sofue et.al(2009).}
\end{figure}
\end{center}
\clearpage
\begin{table}[h]
\begin{center}
\centering
 \label{parameters}
 \begin{tabular}{cccccccccc}
 \hline
 \multicolumn{2}{c}{Source} & &  & \\ \cline{1-2}
   $l^{o}$ & $b^{o}$&$Log_{10}[W_{near}]$& {$N_{He+}/N_{H+}$}& Comments  \\
 & & & &   \\ \hline
 & & & & \\
  7.2&-0.7&2.43&$<$0.018&-\\
 
 15.8&-0.5&2.86&0.064&- \\
 
 17.4&+1.5&2.80&0.082&-\\ 
 
 18.6&-0.8&1.90&$<$0.023&H,He $V_{LSR}$ disagreement \\ 
 
 19.2&+1.7&2.80&0.069&He width broad, but mulitple H components\\
  
 23.4&-0.6&1.76&$<$0.017&Significant pressure broadening\\ 
 
 26.5&0.0&1.91&$<$0.01&-\\ 
 
 28.0&0.0&1.56&$<$0.014&-\\ 
 
 16.3&+1.3&3.16&$<$0.03&-\\
 
 18.0&+1.8&2.76&0.067&Broad He,but $<$ H width, single H component\\
 
 12.6&-0.6&2.74&0.057&-\\ 
 & & & &  \\
 \hline
 \end{tabular}
 \caption{Correlation Weights for line originating ELDWIM towards different positions. $W_{near}$ (Eqn (1))is 
weight for near by ELDWIM clouds. The near and far distances are two possible solutions along the line of sight given 
by the differential Galactic rotation curve. Here only the near distance has been considered, hence $W \rightarrow W_{near}$ 
with $D_c$ being the near cloud distance in Equation(1).}
\end{center}
\end{table}
\vspace{1.8cm}
\begin{center}
\begin{figure}[h]
\vspace{-1.8cm} \hspace{2.0cm}\includegraphics[width=80mm,height=110mm,angle=-90]{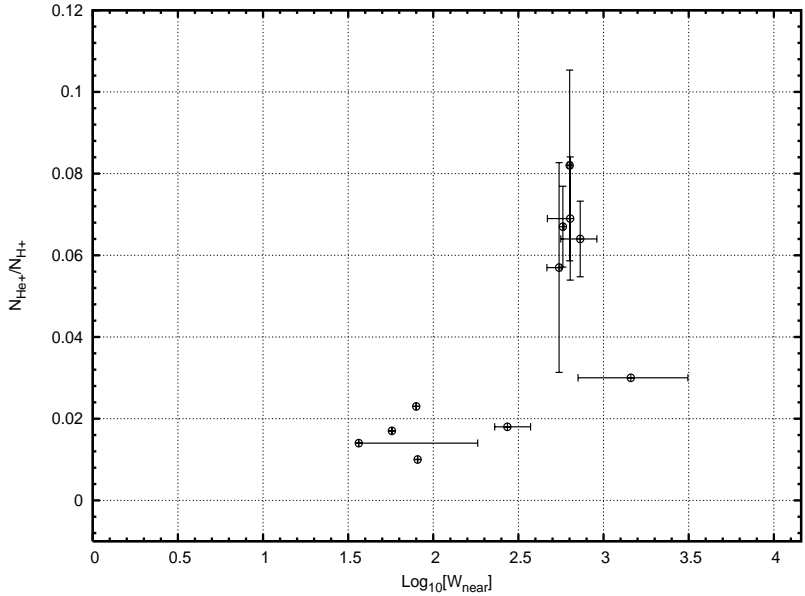}
\caption{Plot of $N_{He+}/N_{H+}$ Vs $Log_{10}\left[W_{near}\right]$ for all the 11 positions.}
\end{figure}
\end{center}
\vspace{-1.0cm}
\section[]{Summary}
This work presents the results of hydrogen and helium RL observations towards 11 different Galactic 
positions using WSRT at $\sim$ 1.4GHz in incoherent addition mode. Out of 11, 5 positions exhibit 
helium RL detection with snr $>$4$\sigma$. WSRT provided 8 IF bands each of width 5 MHz which were used to observe 8 distinct RL 
lines(Hn$\alpha$/Hen$\alpha$ with n=165-171 and H208$\beta$) using frequency switching. Details of observation and data analysis 
have been given in sec 2 \& sec 3. These observations aimed at detecting helium RLs from ELDWIM. Since the abundance(~0.1) 
of helium is smaller than that of hydrogen it is difficult to detect helium RLs compared to the corresponding 
hydrogen RLs. In view of this a set of 15 positions was constructed by consulting previous RL observations
(Lockman et.al 1976, 1989; Heiles et.al 1996a) and the 11 cm continuum map(Reich et al 1990). 
In this procedure a region near to a previous strong($\sim$ 50mK) hydrogen RL detection at 1.4 GHz was 
considered. Next during positioning of the beam on the 11 cm map care was taken not to include any possible 
HII region within the beam. Line detections from 11 of these positions have been displayed in Figure 1 \& 2. 
The fitted gaussian parameters are tabulated in Table-2, along with the obtained ratio $N_{He+}/N_{H+}$. 
Which is the ratio of line 
amplitudes of helium and hydrogen. A detailed discussion of line parameters and their implication has been 
given in sec 4. In general most of the regions seem to be fit to represent ELDWIM. This investigation has 
produced a high value(0.082) of $N_{He+}/N_{H+}$ ratio which is close to the accepted abundance(0.1) of helium to 
hydrogen. Indicating the presence of diffuse ionized helium in ELDWIM. Further a weighting scheme(Eqn (1)) has 
been adopted to show correlation between line orginating ELDWIM and the surrounding HII regions(Paladini et.al 2003 
\& Anderson et.al 2011). According to this scheme a weight is assigned to each region depending on (i)the number of 
surrounding HII regions within a certain radius from the cloud (ii)their distance from the cloud and (iii)the distance(nearest) 
of the cloud from the solar system(Figure 5). This correlation has been displayed by plotting $N_{He+}/N_{H+}$ 
against the obtained weights, $Log_{10}[W_{near}]$. The conclusion from this plot is that regions with high value 
of $N_{He+}/N_{H+}$ bear a high weight. Indicating a correlation between HII regions and helium line detection.        

\section{Acknowledgement} 
The author thanks the staff of WSRT who have made these observations possible. The Westerbork Synthesis Radio Telescope is 
operated by the ASTRON(Netherlands Institute for Radio Astronomy) with support from the Netherlands Foundation for 
Scientific Research(NWO). \\

The author thanks the refree for his comments and suggestions that significantly improved the presentation of this paper.

\label{lastpage}

\end{document}